\documentclass{elsart}

\usepackage{graphicx}
\usepackage{amssymb}
\usepackage{color}
\usepackage{amsmath}
\usepackage{booktabs}
\usepackage[all]{xy}
\newtheorem{theorem}{Theorem}
\newtheorem{proposition}{Proposition}


\makeatletter

\DeclareFontFamily{U}{cyr}{}
\DeclareFontShape{U}{cyr}{m}{n}{
  <5> wncyr5 <6> wncyr6 <7> wncyr7 <8> wncyr8 <9> wncyr9 <10->
wncyr10}{}
\DeclareMathAlphabet{\mathcyr}{U}{cyr}{m}{n}

\input cyracc.def
\input epsf

\newcommand{\ZZ}{{\bf Z}}

\newcommand{\RR}{{\bf R}}
\newcommand{\CC}{{\bf C}}

\newcommand{\NN}{{\bf N}} %\def\NN{{\bf N}}
\newcommand{\FF}{{\bf F}}
\newcommand{\HH}{{\bf H}}

\newcommand{\be}{\begin{equation}}
\newcommand{\ee}{\end{equation}}
\newcommand{\bes}{\begin{equation*}}
\newcommand{\ees}{\end{equation*}}

\DeclareMathOperator{\modular}{Sp}
\DeclareMathOperator{\dimm}{dim}
\DeclareMathOperator{\codim}{codim}
\DeclareMathOperator{\diag}{diag}
\DeclareMathOperator{\sym}{Sym}
\DeclareMathOperator{\Imm}{Im}
\DeclareMathOperator{\Res}{Res}
\DeclareMathOperator{\Ind}{Ind}

\begin{document}

\begin{frontmatter}

\title{More on superstring chiral measures}

\author[insubria]{Francesco Dalla~Piazza}
\ead{f.dallapiazza@uninsubria.it}

\address[insubria]{Dipartimento di Scienze Fisiche e Matematiche, Universit\`a dell'Insubria, 
Via Valleggio 11, I-22100 Como, Italia}  

\begin{abstract}
In this paper we study the expressions of the superstring chiral
measures for $g\leq 5$. We obtain certain new expressions which are
functions of higher powers of theta constants. For $g=3$ we show that
the measures can be written in terms of fourth power of theta constants and for $g=4$ in terms of squares of theta constants. In both
cases the forms $\Xi_8^{(g)}[0^{(g)}]$ appearing in the expression of
the measures are defined on the whole Siegel upper half space. 
Instead, for $g=5$ we find a form $\Xi_8^{(5)}[0^{(5)}]$
which is a polynomial in the classical theta constants, well defined on the
Siegel upper half space and
satisfying some suitable constraints on the moduli space of curves
(and not on the whole Siegel upper half space) that could be a candidate for the
genus five superstring measure. Moreover, we discuss the problem of the
uniqueness of this form in genus five.
%Instead, for $g\geq 5$ we find that the measures can no more be expressed as polynomials in the theta constants and are well defined on the moduli space of curves only.
We also determine the dimension of certain spaces of modular forms and reinterpret the vanishing of the cosmological constant in terms of group representations.
\end{abstract}


\end{frontmatter}

\section{Introduction}\label{intro}
%\subsection{}
String theory in the perturbative approach can be formulated using the
path integral formalism outlined by Polyakov. The starting point is
the functional integral over all the fields in the theory, the
embeddings and the metrics. Employing the symmetries of the theory one
reduces this infinite dimensional integral to a finite dimensional one
over the moduli space of suitable Riemann surfaces. A Riemann surface
can be recovered from its period matrix (see Appendix A). Expanding
such an integral in series over the genus of the Riemann surfaces one can compute perturbatively the scattering amplitudes. This was emphasised in part by Belavin and Knizhnik \cite{belavin_alggeom} who conjectured that ``any multiloop amplitude in any conformal invariant string theory may be
deduced from purely algebraic objects on moduli spaces $M_p$ of Riemann surfaces''. These amplitudes were computed in the 80's up to the four loop order for the bosonic string, and at zero and one loop for the supersymmetric case. In a series of papers D'Hoker and Phong determined an expression for the two loop superstring measure. Recently, in \cite{CDG}, a candidate was proposed for the three loop amplitude. In all these cases the measure is expressed in terms of suitable polynomials in the theta constants.
The latter are defined on the Siegel upper half space $\HH_g$ and not just on the subvariety $J_g\subseteq \HH_g$ of period matrices of genus $g$ Riemann surfaces. This makes the superstring measure, for $g\leq 3$, a function over the whole $\HH_g$.
%This fact, though interesting, is not completely surprising because, for $g\leq 3$, $\HH_g$ and $J_g$ have the same dimension. 
This fact is not completely surprising because, for $g\leq 3$, $\HH_g$ and $J_g$ have the same dimension and $J_g$ is an open set of $\HH_g$. 
In \cite{CDG2} we proposed an expression for the four loop superstring measure which again turned out to be defined over the whole $\HH_4$; this is a remarkable fact, the dimension of the two varieties being no longer the same, see Table \ref{tab:dim}.
\begin{table}[h!]
\begin{center}
\[
\begin{array}{ccccccc}
\toprule
g & &\dimm \HH_g && \dimm J_g && \codim_{\HH_g}J_g \\
\midrule
2 && 3 && 3 && 0 \\
3 && 6 && 6 && 0 \\
4 && 10 && 9 && 1 \\
5 && 15 && 12 && 3\\
g &\phantom{aa}& \frac{1}{2}g(g+1) &\phantom{aa}& 3g-3 &\phantom{aa}& \frac{1}{2}(g-2)(g-3) \\
\bottomrule
\end{array}
\]
\end{center}
\caption{Dimensions of the varieties $\HH_g$ and $J_g$.}
\label{tab:dim}
\end{table}
In \cite{Grr} a candidate was proposed for the superstring measure for any genus $g$. However, these are not a priori well defined for $g\geq 5$ due to the presence of roots. In \cite{SM}, it was proved that for $g=5$ this measure is well defined, at least on $J_g$.
From these facts it is quite natural to investigate if the measure for
$g=5$ could be extended, using the classical theta constants, over all
$\HH_5$ and what happens for $g>5$.
%and how far one can repeat this trick.
Recently in \cite{OPSY} a candidate for the superstring measure for
$g=5$ was
proposed employing the notion of the lattice theta series. This
formalism is almost equivalent to the one of the classical theta constants. Actually, the spaces spanned by theta series and
the ones generated by the bases for the
$O^+$-invariants defined in Sections \ref{f88}, \ref{oinvg4} and \ref{oinvg5} are the same. In genus five both formalisms lead to the same solutions, providing we add to the constraints also the request of the vanishing of the cosmological constant, as we have shown in \cite{CDGd}.

The proposal for the $g$-loop superstring measure rests on the
ansatz (not yet proved) of D'Hoker and Phong \cite{DP1} that the
genus $g$ vacuum to vacuum amplitude takes the form of an integral
over the moduli space of genus $g$ Riemann surfaces of a suitable
differential form that splits into a holomorphic and anti-holomorphic
part. Moreover, the measure $d\mu[\Delta^{(g)}]$ should satisfy
certain reasonable constraints. This characterises it uniquely for
$g\leq 4$ (see below).
In this paper we prove that, assuming these features for the
amplitude, the superstring measure can be defined on the whole $\HH_g$
for $g\leq 5$, but for $g=5$ the correct restriction (see point 3 of
theorem \ref{theo1}) holds true just on $J_4$. This result is stated by the following:
\begin{theorem}\label{theo1}
If the genus $g$ vacuum to vacuum amplitude takes the general form:
\begin{equation}\label{ampl}
\mathcal{A}=\int_{\mathcal{M}_g}(\det\Imm\tau)^{-5}\sum_{\Delta,\bar{\Delta}}c_{\Delta,\bar{\Delta}}d\mu[\Delta^{(g)}](\tau)\wedge
\overline{d\mu[\bar{\Delta^{(g)}}](\tau)},
\end{equation}
where the form $d\mu[\Delta^{(g)}]$ can be written as:
\bes
d\mu[\Delta^{(g)}]=c_g\Xi^{(g)}_8[\Delta^{(g)}](\tau^{(g)})d\mu_B^{(g)}
\ees
and the functions $\Xi^{(g)}_8[\Delta^{(g)}]$ satisfy the following ans\"atze:
\begin{enumerate}
\item they are holomorphic functions on $J_g$;
\item under the action of $\Gamma_g=\modular(2g,\ZZ)$ they should transform as $\Xi^{(g)}_8[M\cdot~\Delta^{(g)}](M\cdot~\tau)\,=\,
\det(C\tau+D)^8\Xi^{(g)}_8[\Delta^{(g)}](\tau),
$
for all $M\in\modular(2g,\ZZ)$;
\item the restriction of these functions to 'reducible' period matrices is a product of the corresponding functions in lower genus;
\end{enumerate}
then $\Xi^{(g)}_8[\Delta^{(g)}]$, and so $d\mu[\Delta^{(g)}]$, are defined everywhere on $\HH_g$, they can be expressed in terms of polynomials
in square of theta constants and they are unique if $g\leq 4$. For $g=5$,
for every even characteristic $\Delta^{(5)}$,
at least one (actually many) forms $\Xi^{(5)}_8[\Delta^{(5)}]$
exist, it is defined on $\HH_5$, it can be written as a polynomial in the theta constants and the restriction requested is
satisfied just on $J_4$. The three constraints do not
characterize it uniquely (at least on $\HH_5$).
\end{theorem}
In the theorem the uniqueness for $g=4$ must be understood as uniqueness
up to a multiple of $J^{(4)}$, that vanishes on the Jacobi
locus\footnote{Note that we indicate, as in literature, with $J^{(g)}$ the modular form
  $J^{(g)}=2^gF_{16}^{(g)}-F_8^{(g)}$, see Section \ref{ginv}, and with
  $J_g$ the Jacobi locus.}$J_4$ (see
below). In genus five, instead, the uniqueness is completely lost.
One
can find many forms $\Xi_8^{(5)}[0^{(5)}]$ that differ not just for
something vanishing on the Jacobi locus. Actually, starting from a form $\Xi_8^{(5)}[0^{(5)}]$ satisfying the three constraints and adding a multiple of $J^{(5)}$, one obtains again a form satisfying the constraints. In genus five $J^{(5)}$ does not vanish on the Jacobi locus, see \cite{GS}. It is still an open problem whether, adding to the constraints the request of the vanishing of the cosmological constant, the uniqueness of the measure is guaranteed. However, the vanishing of the cosmological constant should be automatic for a supersymmetric theory and not imposed by hand. At the moment it is not known if could exist some function satisfying the three constraints and differing from a $\Xi_8^{(5)}$ not just for a multiple of $J^{(5)}$. Moreover, nothing we can say if we consider also the non normal part of the ring of genus five modular forms.

Here $\Delta^{(g)}$ and $\overline{\Delta^{(g)}}$ denote two even genus $g$ theta characteristics, $c_{\Delta,\bar{\Delta}}$ are
suitable constant phases depending on the details of the string model, $d\mu[\Delta^{(g)}](\tau)$ ($\overline{d\mu[\Delta^{(g)}](\tau)}$) is a holomorphic (anti-holomorphic) form
% of maximal rank $(3g-3,0)$ on the moduli space of genus $g$ Riemann surfaces.
and $d\mu_B^{(g)}$ is the well defined genus $g$ bosonic
measure. However, there is not an explicit form for $d\mu_B^{(g)}$ in higher genus.
We observe that the transformation request for the form
$\Xi^{(g)}_8[\Delta^{(g)}]$ is automatic for the integral to make sense, but, as usual, we prefer to emphasize this property for its crucial role in what follows.  

For the genus two and three cases the uniqueness of the forms
$\Xi_8^{(g)}[\Delta^{(g)}]$ is shown in \cite{DG}. In the
same paper, assuming that the
measure is a polynomial in the theta constants, the
uniqueness (up to a term proportional to $J^{(4)}$) is shown for the genus four case and in \cite{OPSY} the general case
is considered (see Section
\ref{smresult} below).
In genus five the uniqueness can not be longer assured. Actually, in
\cite{OPSY}  a candidate for the genus five
superstring measure is proposed. The authors make use of the notion of
the lattice theta
series. An analysis of the different expressions for the chiral
superstring measure can be found in \cite{DbMS, MV}. In a forthcoming paper \cite{CDGd} we will prove that the form
$\Xi_8^{(5)}[0^{(5)}]$ defined there and the one defined in Section
\ref{g5} are different on $\HH_5$ and also on $J_5$. Actually, the difference is  proportional to $J^{(5)}$. The supplementary request of the vanishing of the cosmological constant makes equivalent the two forms.
%The relation between the two
%expressions of the superstring measure
%on the locus of curves is still an open problem.

%It is clear that a similar result holds for $g>5$ because if one could find a polynomial expression in theta constants for the $g>5$ measure, its restriction to $H_1\times H_{g-1}$ would be also polynomial in both factors. This allows to rewrite the non polynomial part of $\Xi^{(5)}_8[0^{(5)}]$ as polynomial in theta constants, but this would be an absurd.

From the result of Salvati Manni \cite{SM}, we know that the square root
appearing in Grushevsky expression of the five loop measure (in the
function $G_5^{(5)}[0^{(5)}]$) is well
defined on the moduli space of curves $J_5$. Further investigations
are needed to understand if, at least, on the locus of curves, it is
polynomial in the classical theta constants.

An indication that the three
constraints cannot define the forms $\Xi_8^{(g)}[\Delta^{(g)}]$
defined over the whole
$\HH_g$ and are sufficient to assure their uniqueness comes from the increasing difference between the dimensions of $\HH_g$ and $J_g$.
The dimension of $\HH_g$ is quadratic in $g$, instead the dimension of
$J_g$ has a linear growth in $g$ and their difference is quadratic in
$g$, see Table \ref{tab:dim}. Thus, it is not surprising that the constraints for the
$\Xi_8^{(g)}[\Delta^{(g)}]$ are not strong enough to characterize it
uniquely.
%Anyway, at the moment, the relations among the restriction
%on $J_g$ of the forms constructed using the theta series satisfying
%the three constraints and the ones defined here
%using the classical theta constants are not known for
%$g=5$.
Further investigations on this point will be carried on in
\cite{CDGd}.

\section{The strategy}\label{strategy}
In \cite{CDG}, inspired by the factorisation of the superstring chiral measure at lower genus, a modification of the an\"satze of D'Hoker and Phong was proposed for the superstring measure. Accordingly, the measure should be written as:
\bes
d\mu[\Delta^{(g)}]=c_g\Xi^{(g)}_8[\Delta^{(g)}](\tau^{(g)})d\mu_B^{(g)},
\ees
where $d\mu_B^{(g)}$ is the bosonic measure at genus $g$ and $\Xi^{(g)}_8[\Delta^{(g)}]$ are suitable functions, see below, $g$ is the genus of the Riemann surfaces considered and $\Delta^{(g)}$ is an even characteristic at genus $g$. The functions $\Xi^{(g)}_8[\Delta^{(g)}]$ are required to satisfy suitable transformation and factorisation constraints, see \cite{CDG} Section 2.2, briefly summarised here:
\begin{enumerate}
\item they must be holomorphic functions on $J_g$;
\item under the action of $\Gamma_g=\modular(2g,\ZZ)$ they should transform as $\Xi^{(g)}_8[M\cdot\Delta^{(g)}](M\cdot\tau)\,=\,
\det(C\tau+D)^8\Xi^{(g)}_8[\Delta^{(g)}](\tau),
$
for all $M\in\modular(2g,\ZZ)$;
\item the restriction of these functions to 'reducible' period matrices is the product of the corresponding functions in lower genus.
\end{enumerate}
Here we emphasise that the first constraint requires that the
forms $\Xi^{(g)}_8[\Delta^{(g)}]$ are defined on the subvariety
$J_g\subset \HH_g$ of period matrices of Riemann surfaces of genus $g$
and not on the whole Siegel upper half space. In fact, $\dimm \HH_g=g(g+1)/2$ and
$\dimm J_g=3g-3$ so these two spaces are the same just for $g\leq
3$. Since we are interested in arbitrary genus, we write $J_g$ instead
of $\HH_g$. Actually, for $g\leq 4 $ the superstring measure can be
extended to the Siegel upper half space, instead for $g=5$ the forms constructed
using the classical theta constants, although well defined over the
whole $\HH_5$, have the correct factorization
just on the Jacobi locus $J_4$, see Section \ref{resg5-14}.

In \cite{CDG} it was pointed out that the $\Xi^{(g)}_8[\Delta^{(g)}]$ are modular forms with respect to the normal subgroup $\Gamma_g(2)$ of $\modular(2g,\ZZ)$ defined by:\begin{align*}
\Gamma_g(2):&=\ker(Sp(2g,\ZZ)\,\longrightarrow\,Sp(2g,\FF_2))\, \\
&=\,
\{M\in Sp(2g,\ZZ):\;A\equiv D\equiv I,\;B\equiv C\equiv 0\;\mbox{mod}\;2\,\},
\end{align*}
where $\FF_2:=\ZZ/2\ZZ$ is the field of two elements. Furthermore we can restrict our attention on a single function
(see, \cite{CDG} Section 2.7) say $\Xi^{(g)}_8[0^{(g)}]$, where $[0^{(g)}]:=[{}_{0\cdots 0}^{0\cdots 0}]$, which has to be a modular form of weight 8 on $\Gamma_g(1,2)$, where $\Gamma_g(1,2)$ is the subgroup of
$\modular(2g,\ZZ)$ which fixes the characteristic $[0^{(g)}]$. In general it is useful to define:
\begin{align*}
\Gamma_g(n,2n):&=\{M\in \Gamma_g(n):\; M\cdot[{}^0_0]\equiv[{}^0_0]\;\mbox{mod}\,2n\}
\, \\
&=\,
\{M\in \Gamma_g:\;{\rm \diag}A{}^tB \equiv {\rm \diag }C{}^tD\equiv \,0\;{\rm mod}\,2n\,\}.
\end{align*}
%Note that if $n$ is even $\Gamma_g(n,2n)$ is a normal subgroup of $\Gamma_g$.
Thus, the $\Xi^{(g)}_8[\Delta^{(g)}]$ are obtained by employing the transitive action of $\modular(2g,\ZZ)$ on the even characteristics. The action of $M\in\modular(2g):=\modular(2g,\FF_2)\cong \Gamma_g/\Gamma_g(2)$ on a characteristic $\Delta^{(g)}$ is given by:
$$
\begin{pmatrix}A&B\\C&D\end{pmatrix}\cdot [{}^a_b]\,:=\,
[{}^c_d],\qquad
\left(\begin{array}{c}{}^tc\\{}^td\end{array}\right)\,=\,
\left(\begin{array}{cc} D&-C\\-B&A\end{array}\right)
\left(\begin{array}{c}{}^ta\\{}^tb\end{array}\right)
\,+\,
\left(\begin{array}{c}{}^t(C\,{}^t\,\!\!D)_0\\{}^t(A\,{}^t\,\!\!B)_0\end{array}\right)
\quad \mbox{mod}\;2,
$$
where $a$, $b$, $c$ and $d$ are the rows of $\Delta^{(g)}$ and $N_0=(N_{11},\ldots,N_{gg})$ is the row vector of diagonal
entries of the matrix $N$.

As explained in \cite{DG}, to which we refer for definitions and notations, the group $\Gamma_g$ acts on the $2^{2g}$ points of $\FF_2^{2g}$ and on the characteristics through its quotient $\modular(2g)\cong \Gamma_g/\Gamma_g(2)$. We defined the subgroup $\Gamma_g(1,2)$ of $\Gamma_g\equiv\Gamma_g(1)$ as the stabiliser of $[0^{(g)}]$ and the image of this subgroup in
$\modular(2g)$ is called $O^+(2g):=\Gamma_g(1,2)/\Gamma_g(2)\subset\modular(2g)$. The three requests, which the function $\Xi^{(g)}_8[0^{(g)}]$ should satisfy, imply that it must belong to the subspace of $O^+$-invariants of weight 8:
$$
M_8(\Gamma_g(2))^{O^+}\,:=\,\{f\in M_8(\Gamma_g(2))\,:\;
\rho(h)f=f\quad \forall h\in O^+(2g)\,\}.
$$
Here $M_k(\Gamma_g(2))$ is the finite dimensional complex vector space of the Siegel modular forms of genus $g$, weight $k$ and level 2 and $\rho$ is the representation of the finite group $\modular(2g)$ on this space defined by:
$$
(\rho(h^{-1})f)(\tau)\,:=\,\det(C\tau+D)^{-k}f(M\cdot \tau),
$$
where $M\in\Gamma_g$ is a representative of $h\in\modular(2g)$ and $f\in M_k(\Gamma_g(2))$.
The action of $\Gamma_g$ on $\tau\in\HH_g$, the Siegel upper half space, is given by:
$$
M\cdot \tau\,:=\,(A\tau+B)(C\tau+D)^{-1},\qquad
M:=\begin{pmatrix}A&B\\C&D\end{pmatrix}\in Sp(2g,\ZZ),\quad\tau\in\HH_g.
$$
Among the functions in $M_8(\Gamma_g(2))$, we will search for the ones
satisfying the three constraints. This is a
general procedure, but for $g>4$ there are some subtleties due to the
loss of the uniqueness of the form $\Xi_8^{(g)}[0^{(g)}]$. Let us briefly summarise these points,
referring to \cite{DG} for more details. To build the $O^+$-invariants we use the $2^g$ second order theta
constants, see \cite{CD2} for details, defined by:
$$
\Theta[\sigma](\tau)\,:=\, \theta[{}^\sigma_{0}](2\tau,0),\qquad
[\sigma]=[\sigma_1\;\sigma_2\;\ldots\;\sigma_g],\;\sigma_i\in\{0,1\},\;
\tau\in\HH_g,
$$
where
$ \theta[{}^a_b](\tau,z)$ are the usual Riemann theta functions. These theta constants are ``modular forms of weight $1/2$'' on $\Gamma_g(2,4)$ and the invariants of degree $4k$ of the quotients group $\Gamma_g(2)/\Gamma_g(2,4)\cong\FF_2^{2g}$ in the ring of polynomials in the $\Theta[\sigma]$'s are modular forms of weight $2k$ on $\Gamma_g(2)$. Indeed, due to
the half integer weight, the group acting on the theta constants is the Heisenberg group, the central extension of
the group  $\Gamma_g(2)/\Gamma_g(2,4)$. Let us denote the space of Heisenberg invariants as
$M_{2k}^{\theta}(\Gamma_g(2))\subset M_{2k}(\Gamma_g(2))$ and
$M_{2k}^\theta(\Gamma_g(2))\,:=\,\CC[\ldots,\Theta[\sigma],\ldots]_{4k}^{H_g}$, where
$\CC[\ldots,\Theta[\sigma],\ldots]_{4k}$ is the subspace of homogeneous polynomials of degree $4k$ in the
$\Theta[\sigma]$'s.
It can be shown (cf.\ \cite{SM2} Thm 2, \cite{R1}, \cite{R2}) that any modular form of weight $2k$ can be written as a homogeneous polynomial of degree $4k$ in the $\Theta[\sigma]$'s if $g\leq 3$, thus
$$
M_{2k}^\theta(\Gamma_g(2))\,=\,M_{2k}(\Gamma_g(2))\qquad
\mbox{for}\quad g=1,2,3.
$$
Such a polynomial is unique for $g<3$, whereas for $g=3$ it is unique
for degree at most than 15 otherwise it is determined
up to the addition of $J^{(3)}G_{4k-16}$ where $G_{4k-16}$ is any homogeneous polynomial of degree $4k-16$ in the $\Theta[\sigma]$'s and $J^{(3)}$ is a polynomial of degree 16 vanishing identically on the $\Theta[\sigma]$'s, see
\cite{vGvdG}. In all other genera there could exist some modular forms which cannot be written as polynomials in
the $\Theta[\sigma]$'s\footnote{In general the graded ring of modular forms of even weight on $\Gamma_g(2)$ is the normalization of the ring of
the $\Theta[\sigma]$'s:
$
\oplus_{k=0}^\infty\,M_{2k}(\Gamma_g(2))\;=\;
(\CC[\ldots,\Theta[\sigma],\ldots]^{H_g})^{Nor}
$.}.

The dimension of the space of $O^+$-invariants can be determined from the decomposition of the
$\modular(2g)$-representation into irreducible representations and using the Frobenius reciprocity. Thus, the dimension
is given by the multiplicity of the trivial representation ${\bf 1}$ of $O^+$ in the $O^+-$representation
$\Res_{O^+}^{\modular(2g)}(V)$:
$$
\dim V^{O^+}\,=\,\langle\,\Res^{Sp}_{O^+}(V),{\bf 1}\,\rangle_{O^+}\,=\,
\langle\,V,\Ind^{Sp}_{O^+}({\bf 1})\,\rangle_{Sp}.
$$
Here $\Res^{Sp}_{O^+}(V)$ is the restriction of the representation
from $Sp(2g)$ to $O^+(2g)$, $\Ind^{Sp}_{O^+}({\bf 1})$ is the induced
representation of the representation ${\bf 1}$ of $O^+(2g)$ to the
whole $Sp(2g)$ and the second identity is the Frobenius identity, see
\cite{DG, CD2, sym}.
Frame \cite{F-ind} showed that $\Ind^{Sp}_{O^+}({\bf 1})=\,{\bf 1}\,+\,\sigma_\theta$, where {\bf 1} is the trivial
representation and $\sigma_\theta$ is an irreducible representation of dimension $2^{g-1}(2^{g}+1)-1$, so that if
the multiplicities of ${\bf 1}$ and $\sigma_\theta$ in $V$ are $n_1$ and $n_{\sigma_\theta}$ respectively, the dimension
of the space of $O^+$-invariants is $\dim V^{O^+}=n_1+n_{\sigma_\theta}$.
\begin{table}[h!]
\begin{center}
\[
\begin{array}{ccccc}
\toprule
g && \sigma_\theta && \dim(\sigma_\theta) \\
\midrule
1 && \rho_{[21]} && 2 \\
2 && \rho_{[42]} && 9 \\
3 && {\bf 35}_b && 35 \\
4 && {\bf 135} && 135 \\
5 &\phantom{aa}& {\bf 527} &\phantom{aa}& 527 \\
\bottomrule
\end{array}
\]
\end{center}
\caption{The $\sigma_\theta$ representations for the low genus cases and their dimensions.}
\label{tab:repsigma}
\end{table}

In what follows we will label the irreducible representations of $\modular(2g)$ with the partitions of 3 and 6 for genus
one and two respectively (recall that $\modular(2) \cong S_3$ and $\modular(4)\cong S_6$), as in \cite{CD}; we will
follow Frame's notation \cite{Frame} for genus three and indicate them just with their dimensions\footnote{If they are not unique at the given size, we will indicate also the character of the second
conjugacy class, the one of the non zero transvections (which has 255 and 1023 elements for genus four and
five respectively). Transvections are analogous to reflections in orthogonal groups (cf.\ \cite{Jac}, $\S$ 6.9 or \cite{DG}).} for
$g\geq 4$.
In Table \ref{tab:repsigma} are reported the $\sigma_\theta$ representations for the lower genus cases. In case $g=1$,
$\rho[21]$ is the unique two dimensional representation of $S_3 \cong \modular(2)$ and $[21]$ is the partition of
3 labelling it. For $g=2$, $\rho[42]$, or $n_9$ in the notations of
\cite{CD}, is the nine dimensional representation of $S_6 \cong
\modular(4)$ for which the character of ${\bf 1}+\sigma_\theta$ is
positive ($[42]$ is the partition of six labelling this irreducible
representation; see \cite{DG} Section 4.2 and \cite {CD2} Section 5.2.1
for the explanation of why the character of the representation must be
positive). For $g=3$, the ${\bf 35}_b$ is
the unique 35 dimensional representation of $\modular(6)$, as reported in \cite{Frame} or as can be computed using, for example, the software Magma.
For $g=4$, ${\bf 135}$ is the unique 135 dimensional irreducible representation of $\modular(8)$ and for $g=5$,
${\bf 527}$ is the unique 527 dimensional irreducible representation of $\modular(10)$, as can be computed using Magma.

In \cite{DG}, the representations of $\modular(2g)$ on
$M_k(\Gamma_g(2))$, the vector space of Siegel modular forms of given weight for
the principal congruence subgroup of level two, were studied for small genus $g$, and decomposed into irreducible
representations.
For the applications in string theory, we are interested in the representations of $\modular(2g)$ on the space
$M_8(\Gamma_g(2))$, the modular forms of weight eight with respect to the group $\Gamma_g(2)$.
Let us report here the decomposition of these representations for $g\leq 3$:
\begin{align*}
M_8(\Gamma_1(2)) & \cong \mathrm{\sym}^4(\rho_{[21]})={\bf 1}+2\rho_{[21]}, \\
M_8(\Gamma_2(2)) & \cong \mathrm{\sym}^4(\rho_{[2^3]})-{\bf 1}={\bf 1}+3\rho_{[2^3]}+3\rho_{[42]}+\rho_{[31^3]}+\rho_{[321]}, \\
M_8(\Gamma_3(2)) &
={\bf 1}+ 4\cdot{\bf 15}_a+{\bf 35}_a+4\cdot{\bf 35}_b+
5\cdot {\bf 84}_a+2\cdot{\bf 105}_c+{\bf 168}_a+\\
&2\cdot{\bf 189}_c+3\cdot{\bf 216}+
3\cdot{\bf 280}_b+2\cdot{\bf 336}_a+{\bf 420}_a.
\end{align*}
From the Frobenius identity it follows that for $g=1$ the dimension of the space of $O^+$-invariants is three, for $g=2$
is four and for $g=3$ is five. For genus four and five we do not know the decomposition of the whole $M_8(\Gamma_g(2))$
and, moreover, the ring of modular forms is not understood in terms of Heisenberg invariant polynomials in theta
constants. However, in Section \ref{g4} and \ref{g5}, we will restrict our attention to the space
$M_8^\theta(\Gamma_g(2))$, searching the $O^+$-invariants there.

\section{The construction of the $O^+$-invariants}
Once the dimension of $M_8(\Gamma_g(2))^{O^+}$ is known, the main
problem is to find an explicit expression for a basis of this space in
terms of theta constants, if possible. In \cite{CDG} and \cite{CDG2}
the notion of isotropic subspaces was employed 
to find such bases. Recall that, if $V$ is provided with a symplectic form, $W\subset V$ is an isotropic subspace if
on every pair of vectors in $W$ the symplectic
form vanishes. This way to determine the invariants makes use of the geometry underlying the theta characteristics and
the corresponding action of the symplectic group on them. For example, the condition for a subspace to be isotropic
is preserved under the action of $\modular(2g)$. Moreover, it is quite simple to determine the restriction on a block
diagonal period matrix of the $O^+$-invariant built in this way, despite to the huge number of terms appearing in
these functions, cf. the discussion in Appendix C of \cite{CDG}.
The knowledge of a basis for these spaces allows to find, for $g\leq 5$, a linear combination of the $O^+$-invariants
such that its restrictions fits all requests in the ans\"atze discussed in Section \ref{strategy}.
%As explained before, for $g=1,2,3$ any modular form of weight $2k$ can be written as a homogeneous polynomial of degree
%$4k$ in the theta constants and this polynomial is unique. This allows to prove
%the uniqueness for the expression of the superstring measure.
For $g=1,2,3$ the fact that any modular form of weight $2k$ can be
expressed as a polynomial of degree $4k$ in theta constants in a unique
way (unique up to a multiple of $J^{(3)}$ if $g=3$ and $k>4$, see
Section \ref{strategy}) allows
us to prove the uniqueness for the expression of the superstring measure.
In genus four the ring of Siegel modular forms is not
normal. This means that in general there could be some modular forms that cannot be expressed as polynomials in
theta constants. In this case in \cite{DG} the uniqueness was proved in a weakened form, assuming the polynomiality for
the amplitude, i.e. considering $O^+$-invariants contained in the
space $M_8^\theta(\Gamma_4(2))$ only. In \cite{OPSY} the proof is
also extended to the general case. As anticipated in Section
\ref{intro}, in genus five the three constraints are not strong enough
to assure the uniqueness of the superstring measure neither if we
restrict to the normal part of the ring of modular forms as we will
prove in Section \ref{g5} and in \cite{CDGd}. Thus, in genus five the loss of the uniqueness
is not due just to the non normality of the ring of modular forms.

In \cite{Grr} a generalisation for the expression of the chiral measure at any genus $g$ was proposed. In the
approach used there, the action of the symplectic group underlying that expression is not manifest although the correct
factorisation is obtained.
The author restricts the search for the $g$ loop amplitudes to a suitable vector space of dimension $g+1$ then finding
there a unique solution of the constraints. However, for genus three (four and more) the vector space defined by
the transformation constraint has dimension five ($\geq 7$), see
\cite{DG} 7.4 and 7.5, which is larger than the dimensions of the starting spaces
selected in \cite{Grr}.
Moreover, his expression might be not well-defined for $g>5$ (Salvati Manni in \cite{SM} discusses the case $g=5$) due to
the presence of some roots. 

We will now provide new expressions for $\Xi^{(g)}_8[\Delta^{(g)}]$ at
lower genus. In these new formulas, the theta constants appear at
higher power than in the expressions given in all previous works.

\subsection{$O^+$-invariants for the genus 3 case}\label{f88}
For genus $g=3$ the only $\modular(6)$-representations that have an $O^+$-invariant are ${\bf 1}$ and
$\sigma_\theta={\bf 35}_b$ and we know that there are five such linearly independent invariants, see \cite{DG}.
The representation ${\bf 1}$ provides the $\modular(6)$-invariant and ${\bf 35}_b$ the representation on
the $\theta[\Delta^{(3)}]^8$. A natural question is about the number of linearly independent $O^+$-invariants of degree
16 that can be written as quadratic polynomials in $\theta[\Delta^{(3)}]^8$.
From the decomposition of the tensor products in irreducible representations we find that:
\begin{align*}
\sym^2({\bf 1}+{\bf 35}_b)&={\bf 1}+{\bf 35}_b+\sym^2({\bf 35}_b) \\
&={\bf 1}+{\bf 35}_b+{\bf 1}+{\bf 27}_a+2\cdot{\bf 35}_b+{\bf 84}_a+{\bf 168}_a+{\bf 280}_b\\
&=2\cdot{\bf 1}+{\bf 27}_a+3\cdot{\bf 35}_b+{\bf 84}_a+{\bf 168}_a+{\bf 280}_b,
\end{align*}
so that we get the two $\modular(6)$-invariants $\sum_\Delta\theta[\Delta^{(3)}]^{16}$ and
$(\sum_\Delta\theta[\Delta^{(3)}]^{8})^2$ and three $O^+$-invariants (but not $\modular(6)$), two of which are
$\theta[0^{(3)}]^{16}$ and $\theta[0^{(3)}]^8\sum_\Delta\theta[\Delta^{(3)}]^8$.
In order to find the third invariant quadratic in
$\theta[\Delta^{(3)}]^8$ we can adopt a general method that allows us
to generate many $O^+$-invariants (clearly not all independent). This consists in starting from  a certain monomial
of degree sixteen which contains the theta constants to the power at least four, and imposing some suitable condition
on the corresponding characteristics. For example, in the spirit of \cite{DP1} and \cite{DP2}, we can take
$\theta[\Delta_1^{(3)}]^4\theta[\Delta_2^{(3)}]^4\theta[\Delta_3^{(3)}]^4\theta[\Delta_4^{(3)}]^4$
with the conditions $\Delta_1+\Delta_2+\Delta_3+\Delta_4=0^{(3)}$.
There are 1611 such monomials which summed up give an $O^+$-invariant. In fact, this is redundant because there
are ``sub-polynomials'' that are orbits for $O^+$ and then are themselves invariant.
In \cite{DG} Section 8.5, the generators of $O^+\cong S_8$ are given in terms of transvections acting on the theta constants.
Thus, an orbit can be determined acting on a single monomial with these transvections until the number of terms of
the generated polynomial stops to grow. Next, one considers a second
monomial (not in the orbit of the first)
and repeats the procedure.
In this way we can recognise eight $O^+$-invariants inside the big polynomial, as shown in Table \ref{tab:orbits}.
\begin{table}[h!]
\begin{center}
\[
\begin{array}{cccp{4.5cm}c}
\toprule
\mbox{orbit} & \mbox{general expression} && \mbox{condition} & \mbox{num. elem.} \\
\midrule
1 & \theta[0]^{16} && &1 \\
2 & \theta[\Delta]^{16} && $\Delta\neq 0$ & 35 \\
3 & \theta[0]^{8}\sum_\Delta\theta[\Delta]^8 && $\Delta\neq 0$ & 35 \\
4 & \theta[0]^{4}\sum_{\Delta_i,\Delta_j,\Delta_k}\theta[\Delta_i]^4\theta[\Delta_j]^{4}\theta[\Delta_k]^4 && $\Delta_i+\Delta_j+\Delta_k=0$ \newline $\Delta_i,\Delta_j,\Delta_k\neq 0$ \newline $\Delta_i\neq\Delta_j\neq\Delta_k$ & 105 \\
5 & \sum_{\Delta_i,\Delta_j,\Delta_k,\Delta_l}\theta[\Delta_i]^4\theta[\Delta_j]^{4}\theta[\Delta_k]^4\theta[\Delta_l]^4 && $\Delta_i+\Delta_j+\Delta_k+\Delta_l=0$ \newline $\Delta_i+\Delta_j+\Delta_k$ even \newline $\Delta_i+\Delta_j$ even \newline $\Delta_i,\Delta_j,\Delta_k,\Delta_l\neq 0$ \newline $\Delta_i\neq\Delta_j\neq\Delta_k\neq\Delta_l $ & 210 \\
6 &\sum_{\Delta_i,\Delta_j}\theta[\Delta_i]^8\theta[\Delta_j]^8 && $\Delta_i+\Delta_j$ odd \newline $\Delta_i,\Delta_j\neq 0$ \newline $\Delta_i\neq\Delta_j$ & 280 \\
7 &\sum_{\Delta_i,\Delta_j}\theta[\Delta_i]^8\theta[\Delta_j]^8 & & $\Delta_i+\Delta_j$ even \newline $\Delta_i,\Delta_j\neq 0$ \newline $\Delta_i\neq\Delta_j$ & 315 \\
8 & \sum_{\Delta_i,\Delta_j,\Delta_k,\Delta_l}\theta[\Delta_i]^4\theta[\Delta_j]^{4}\theta[\Delta_k]^4\theta[\Delta_l]^4 &\phantom{aa}& $\Delta_i+\Delta_j+\Delta_k+\Delta_l=0$ \newline $\Delta_i+\Delta_j+\Delta_k$ even \newline $\Delta_i+\Delta_j$ even \newline $\Delta_k+\Delta_l$ odd \newline $\Delta_i,\Delta_j,\Delta_k,\Delta_l\neq 0$ \newline $\Delta_i\neq\Delta_j\neq\Delta_k\neq\Delta_l $ & 630 \\
\bottomrule
\end{array}
\]
\end{center}
\caption{Orbits under the action of $O^+$ (genus three case).}
\label{tab:orbits}
\end{table}
It is clear that the searched invariant could be the sixth or the
seventh. Using a computer or (quite lengthy!) by hand and the
classical theta formula (cf. \cite{CDG}, Section 3.2), we verify that each of them is linearly independent from the other invariants. We then choose
the sixth, which we will call $F_{88}^{(3)}$ (and $F_{88}^{(g)}$ for arbitrary genus $g$).
Thus, each monomial in $F_{88}^{(3)}$ is the product of two theta constants at the eighth power, with the
conditions that the sum of their characteristics is odd (even, if we choose the seventh), the two characteristics are
not equal and both are not zero.
Note the following equality between the $O^+$-invariants:
\begin{align*}
\left(\sum_\Delta \theta[\Delta^{(3)}]^8\right)^2-\sum_\Delta\theta[\Delta^{(3)}]^{16} &=2\sum_{(\Delta_i,\Delta_j)_e}\theta[\Delta_i^{(3)}]^8\theta[\Delta_j^{(3)}]^8 +2\sum_{(\Delta_i,\Delta_j)_o}\theta[\Delta_i^{(3)}]^8\theta[\Delta_j^{(3)}]^8\\
&+2(\theta[0^{(g)}]^8\sum_{\Delta}\theta[\Delta^{(3)}]^8-\theta[0^{(3)}]^{16}),
\end{align*}
where the first two functions on the r.h.s. are the $O^+$-invariants of lines six and seven of the Table \ref{tab:orbits} and
the ``e'' and ``o'' stand for even sum and odd sum of the two
characteristics respectively. The two functions on the l.h.s. are the
two $\modular(6)$-invariants, $F_8^{(3)}$ and $F_{16}^{(3)}$ as we will
call them in the following.
In fact, for genus three the two $\modular(6)$-invariants are not linearly independent but there is a relation between
them (the $J^{(3)}$, see below, or $F_{16}$ in the notation of \cite{DG}). Therefore, to find a basis we need to look for
another invariant which cannot be expressed as a quadratic polynomial in $\theta[\Delta^{(3)}]^8$. We can take
$\theta[0^{(3)}]^4\sum_\Delta\theta[\Delta^{(3)}]^{12}$.

In Section \ref{g3} we will show how to build the chiral measure from these functions.

\subsection{$O^+$-invariants for the genus 4 case.}\label{oinvg4}
For genus $g=4$, the only $\modular(8)$-representations containing an $O^+$-invariant are ${\bf 1}$ and ${\bf 135}$.
Now, it is not known if $M_{2k}^\theta(\Gamma_4(2))$, the space of modular forms of weight $2k$ which are
(Heisenberg-invariants) polynomial in $\Theta[\sigma]$'s, coincides with $M_{2k}(\Gamma_4(2))$.
Recently, Oura determined the dimension of $M_{2k}^\theta(\Gamma_g(2))^{O^+}$ obtaining 7 for the $g=4$ case. In
principle these dimensions could also be computed using a method similar to those we used for $g<4$, i.e. searching for
the decomposition of $M_8^{\theta}(\Gamma_g(2))$ in irreducible representations, but it is very time and memory
consuming for increasing $g$.
As for the genus three case, we want to find a basis for the $O^+$-invariants in which the theta constants appear with
the highest possible degree. Let us start by determining the decomposition of the symmetric product
$\sym^2({\bf 1}+{\bf 135})$ in irreducible representations. This can be done using Magma or, by hand, with the
character table of $\modular(8)$ and the character inner product (see
\cite{CD2}, Section 5.2.1,  for the case $g=2$).
We obtain the decomposition:
\begin{align*}
\sym^2({\bf 1}+{\bf 135})&={\bf 1}+{\bf 135}+\sym^2({\bf 135}) \\
&=2\cdot{\bf 1}+{\bf 119}+3\cdot{\bf 135}+{\bf 1190}+{\bf 3400}+{\bf 4200}.
\end{align*}
This means that we can find five $O^+$-invariants that are quadratic polynomials in the theta constants at the
eighth power. Two of them are the $\modular(8)$-invariants $\sum_\Delta\theta[\Delta^{(4)}]^{16}$ and
$(\sum_\Delta\theta[\Delta^{(4)}]^{8})^2$ which are now linearly independent because the Schottky relation $J^{(4)}$, the analogous of $J^{(3)}$ for genus four (see Section \ref{ginv} for the definition),
vanishes just on $J_4$ and not identically on the whole $\HH_4$. The remaining three invariants are $\theta[0^{(4)}]^{16}$,
$\theta[0^{(4)}]^8\sum_\Delta\theta[\Delta^{(4)}]^8$ and the generalisation of the $O^+$-invariant found in Section
\ref{f88} to the genus four case, $F_{88}^{(4)}$ (the construction of
such a function for $g\geq4$ is straightforward).

We now check how many $O^+$-invariants can be written as polynomials of degree four in the $\theta[\Delta^{(4)}]^4$.
This can be done decomposing the symmetric product $\sym^4(\rho_\theta)$ in irreducible representations\footnote{Here
$\rho_\theta$ is the representation of $\modular(2g)$ on the $\theta[\Delta^{(g)}]^4$ ($2g=8$ in this case).}, and
counting the multiplicity of the representations ${\bf 1}$ and $\sigma_\theta$ ($\sigma_\theta={\bf 135}$ in this case).
In \cite{vG} it was shown that the representation of $\modular(2g)$ on the subspace
$M_2^\theta(\Gamma_g(2))\subset M_2(\Gamma_g(2))$, that is spanned by the $\theta[\Delta^{(g)}]^4$, is isomorphic to
the representation $\rho_\theta$ found by Frame \cite{F-ind} that supports $O^+$-anti-invariants. This representation
has dimension $\dim \rho_\theta=(2^g+1)(2^{g-1}+1)/3$, so for $g=4$ one finds $\rho_\theta={\bf 51}$; see \cite{DG}
for details. Thus, a function belonging to $\sym^{2n}(\rho_\theta)$ is an $O^+$-invariant of degree $2n$ in
$\theta[\Delta^{(g)}]^4$, $n\in\NN$. We have:
\begin{align*}
\sym^4({\bf 51})&=2\cdot{\bf 1}+{\bf 51}+{\bf 119}+4\cdot{\bf 135}+{\bf 510}+2\cdot{\bf 918}+5\cdot{\bf 1190}+{\bf 1275}\\
&+{\bf 2856}_{-504}+2\cdot{\bf 3400}+3\cdot{\bf 4200}+{\bf 5712}+{\bf 5950}_{-210}+{\bf 7140}\\
&+{\bf 8160}+{\bf 11900}_{700}+3\cdot{\bf 13600}+{\bf 18360}+2\cdot{\bf 19040}+{\bf 23800}_{-1960}\\
&+{\bf 32130}_{2898}+{\bf 34560}+{\bf 57120},
\end{align*}
so we get six $O^{+}$-invariants, the five found before and $\theta[0^{(4)}]^4\sum_{\Delta}\theta[\Delta^{(4)}]^{12}$.
The seventh invariant cannot be written in this way, but we can search
for it as a polynomial in the $\theta[\Delta^{(2)}]^2$.
In general, we have not a representation of $\modular(2g)$ on the space generated by the $\theta[\Delta^{(g)}]^2$. So we
cannot repeat the previous method using something like $\sym^8(\cdots)$. However, we already know at
least one $O^+$-invariant linearly independent from the others that can be written as a polynomial in
$\theta[\Delta^{(4)}]^2$: the invariant $G_1[0^{(4)}]$ defined  in \cite{CDG2}, which, for later convenience, will
be renamed\footnote{We made a change of notation with respect our previous works: all the forms
built using the isotropic space will be indicated by $G_d^{(g)}[0^{(g)}]$, where $d$ is the dimension of the isotropic subspace
and $g$ the genus we are considering. For example the form $H[0^{(3)}]$ of \cite{CDG2} becomes $G_2^{(3)}[0^{(3)}]$ in the
new notation.} $G_3^{(4)}[0^{(4)}]$. Using a computer, we have
verified that it is linearly independent from the other six. The same
conclusion can be achieved using the approach of \cite{CDGd}.
Let us recall the definition of the function $G_3^{(4)}[0^{(4)}]$, or $P_{3,2}^{(4)}$ in Grushevsky notation
\cite{Grr}.
Given any three dimensional isotropic subspace $W\in \FF_2^8$,
there are $3\cdot 8=24$ even quadrics $Q_\Delta$ such that
$W\subset Q_\Delta$. Let $Q_0\subset \FF_2^8$ be the even quadric
with characteristic $\Delta_0^{(4)}=[0^{(4)}]$.
%Moreover, if $W\subset Q_\Delta$ then also $W\subset Q_{\Delta+w}$
%for all $w\in W$.
We will use only the octets of quadrics which contain $Q_0$ to define a modular form $G_3^{(4)}[0^{(4)}]$:
$$
G_3^{(4)}[0^{(4)}]\,=\,\sum_{W\subset Q_0}\,\,\prod_{w\in W}\,
\theta[\Delta_0^{(4)}+w]^2,
$$
where we sum over the $2025$ three dimensional isotropic subspaces
$W\subset Q_0$,
and for each such subspace we take the product of
the eight even $\theta[\Delta_0^{(4)}+w]^2$. As explained in \cite{CDG2} it is a modular form on $\Gamma_4(1,2)$.

Having found seven linearly independent $O^+$-invariants, according to Oura's result, we have a basis for
$M_8^\theta(\Gamma_4(2))^{O^+}$ and in Section \ref{g4} we will search for a linear combination of them to build
the function $\Xi^{(4)}_8[0^{(4)}]$ which restricts correctly.

\subsection{$O^+$-invariants for the genus 5 case}\label{oinvg5}
For genus five, the ring of modular forms, as for $g=4$, is not normal. Moreover, there may exist many relations satisfied
by the second order theta constants, but in any case they are not known. Finally, the Schottky
relation does not vanish on the Jacobi locus (although this was
conjectured by Belavin and Knizhnik \cite{bk}, Conjecture 3, by
Morozov and Perelemov \cite{bkmp,Mo3} and by D'Hoker and Phong
in \cite{DP2}, Section 4.1 and it was shown that it vanishes for any
genus on the
hyperelliptic locus by Poor \cite{P}): a very recent result \cite{GS}, Corollary 18, shows that the zero locus of this form is the locus of trigonal curves.

Despite these difficulties, starting from the seven functions and mimicking the $g=4$ invariants, we can try to add a
further linearly independent $O^+$-invariant polynomial and look for a linear combination (possibly unique) which
factorises in the right way.
Indeed, for genus five it is known that
\begin{equation}\label{inequal}
%\dim [\Gamma_5(1,2),8]^{\theta^2} \leq \dim
%[\Gamma_5(1,2),8]^{\theta} \leq \dim [\Gamma_5(1,2),8]^{\theta_S}=8,
\dim M_8 ^{\theta^2} (\Gamma_5(2))^{O^+} \leq \dim M_8 ^{\theta} (\Gamma_5(2))^{O^+} \leq \dim M_8 ^{\theta_S} (\Gamma_5(2))^{O^+}=8,
\end{equation}
where the first term is the space of modular forms with respect to the group $\Gamma_5(1,2)$ of weight eight which are polynomial in $\theta[\Delta^{(5)}]^2$,
the second is the space of modular forms polynomial in $\theta[\Delta^{(5)}]$ (w.r.t. the same group and of same weight as before), and the third is the space of the theta
series associated to quadratic forms, see \cite{andr}. Note that it is
not clear that
$M_8^{\theta}(\Gamma_5(2))^{O^+}$ is a subset of
$M_8^{\theta_S}(\Gamma_5(2))^{O^+}$, in fact in paper \cite{CDGd}
we have shown that these two spaces are the same.
%$[\Gamma_5(1,2),8]^{\theta} \subseteq [\Gamma_5(1,2),8]^{\theta_S}$.

In order to construct a basis for $M_8^\theta(\Gamma_5(2))^{O^+}$ we generalise the form $G_3^{(4)}[0^{(4)}]$ used
before and the inequalities \eqref{inequal} show that we can find at most one $O^+$-invariant polynomial in
$\theta[\Delta^{(5)}]$. To this aim we consider the form $G_4^{(4)}[0^{(4)}]$ introduced in \cite{CDG2} and define
it also for the $g=5$ case. This goes straightforward with the notion of isotropic subspaces and in principle we
can define similar forms for arbitrary genus $g$ using isotropic subspaces of dimension at most $g$ (see e.g.\ \cite{Grr}).

\subsubsection{The form $G_3^{(5)}[0^{(5)}]$}
We will follow the definitions of \cite{CDG2}.
Let $W\subset\FF_2^{10}$ be a three dimensional isotropic subspace. Given such a $W$, there are $10\cdot8=80$ even
quadrics $Q_\Delta$ such that $W\subset Q_\Delta$. Let $Q_0\subset \FF_2^{10}$ be the even quadric with characteristic
$\Delta_0^{(5)}=[0^{(5)}]$. We will only use the octets of quadrics which contain $Q_0$ to define the modular form
$G_3^{(5)}[0^{(5)}]$, or $P_{3,2}^{(5)}$ in the notations of \cite{Grr}:
\bes
G_3^{(5)}[0^{(5)}]\,:=\,\sum_{W\subset Q_0}\,\,\prod_{w\in W}\,
\theta[\Delta_0^{(5)}+w]^2,
\ees
where we sum over the 118575 three dimensional isotropic subspaces $W\subset Q_0$, and for each such subspace we take the product of the
eight $\theta[\Delta_0^{(5)}+w]^2$.
\subsubsection{The form $G_4^{(5)}[0^{(5)}]$}
Let $W\subset\FF_2^{10}$ be a four dimensional isotropic subspace. Given such a $W$, there are 48 even quadrics
$Q_\Delta$ such that $W\subset Q_\Delta$. Let $Q_0\subset \FF_2^{10}$ be the even quadric with characteristic
$\Delta_0^{(5)}=[0^{(5)}]$. We will only use the sets of quadrics which contain $Q_0$ to define the modular
form $G_4^{(5)}[0^{(5)}]$, or $P_{4,1}^{(5)}$ as in \cite{Grr}:
\bes
G_4^{(5)}[0^{(5)}]\,:=\,\sum_{W\subset Q_0}\,\,\prod_{w\in W}\,
\theta[\Delta_0^{(5)}+w],
\ees
where we sum over the 71145 four dimensional isotropic subspaces $W\subset Q_0$, and for each such subspace we take the product of
the sixteen $\theta[\Delta_0^{(5)}+w]$.

\subsubsection{Remark} \label{eightfuncts}
The eight functions $F_1^{(5)}$, $F_2^{(5)}$, $F_3^{(5)}$, $F_8^{(5)}$, $F_{18}^{(5)}$, $F_{16}^{(5)}$, $G_3^{(5)}[0^{(5)}]$, $G_4^{(5)}[0^{(5)}]$ are linearly independent, as can be checked by a computer or by the restriction we will deduce in
Section \ref{g5} or using the technique of \cite{CDGd}. Thus, it follows that in \eqref{inequal} an equality must hold between the second and the third term:
\bes
\dim M_8 ^{\theta} (\Gamma_5(2))^{O^+} = \dim M_8 ^{\theta_S} (\Gamma_5(2))^{O^+}=8
%\dim [\Gamma_5(1,2),8]^{\theta}=\dim [\Gamma_5(1,2),8]^{\theta_S}=8.
\ees
From the computations at genus four, given in Section \ref{g4}, and from a result of Nebe \cite{Ng}, it also follows that at genus five:
\begin{equation} \label{equalg5}
\dim M_8 ^{\theta^2} (\Gamma_5(2))^{O^+}=7.
\end{equation}
Indeed, we obtain seven linear independent functions in the space 
%$[\Gamma_5(1,2),8]^{\theta^2}$ 
$M_8 ^{\theta^2} (\Gamma_5(2))^{O^+}$
so its dimension
is greater or equal than seven. In \cite{Ng} it was determined seven
as an upper limit for the dimension of the space of theta square
series associated to quadratic forms. As this space contains
$M_8 ^{\theta^2} (\Gamma_5(2))^{O^+}$, it follows the equality
\eqref{equalg5}. This result can also be checked using the method of \cite{CDGd}.
This fixes all the dimensions of the spaces appearing in the previous inequality \eqref{inequal}.

\subsection{Genus $g$ expressions for $O^+$-invariants }\label{ginv}
In this Section we recall the six $O^+$-invariants belonging to $\sym^4\rho_\theta$, found for the lower genus,
and we generalise them for arbitrary $g$. The first three are the same as in \cite{CDG}, but multiplied by
$\theta[0^{(g)}]^4$ to get a form of weight eight; $F_8^{(g)}$ and $F_{16}^{(g)}$ are the two $\modular(2g)$-invariants,
$F_{88}^{(g)}$ is the generalised $O^+$-invariant introduced in
Section \ref{f88} and the modular form $J^{(g)}:=2^gF_{16}^{(g)}-F_8^{(g)}$ vanishes identically for genus three, as explained in \cite{CDG}. These are:
\begin{align*}
F_1^{(g)} & :=\,\theta[0^{(g)}]^{16}, \\
F_2^{(g)} & :=\,\theta[0^{(g)}]^4\sum_{\Delta^{(g)}} \,\theta[\Delta^{(g)}]^{12}, \\
F_3^{(g)} & :=\,\theta[0^{(g)}]^8\sum_{\Delta^{(g)}}\theta[\Delta^{(g)}]^8, \\
F_8^{(g)} & :=\,(\sum_{\Delta^{(g)}}\theta[\Delta^{(g)}]^8)^2, \\
F_{88}^{(g)} & :=\,\sum_{(\Delta_i^{(g)},\Delta_j^{(g)})_o}\theta[\Delta_i^{(g)}]^8\theta[\Delta_j^{(g)}]^8,\\
F_{16}^{(g)} & :=\, \sum_{\Delta^{(g)}}\theta[\Delta^{(g)}]^{16}, \\
J^{(g)} & :=\, 2^g\sum_{\Delta^{(g)}}\theta[\Delta^{(g)}]^{16}-(\sum_{\Delta^{(g)}}\theta[\Delta^{(g)}]^8)^2=2^gF_{16}^{(g)}-F_8^{(g)},
\end{align*}
where $(\Delta_i^{(g)},\Delta_j^{(g)})_o$ stands for all the pairs of distinct even characteristics such that their sum
is odd.
Behind these, we also introduced the forms $G_3^{(g)}[0^{(g)}]$ for $g=4,5$ and $G_4^{(g)}[0^{(g)}]$ for $g=5$. However $G_3^{(g)}[0^{(g)}]$
($G_4^{(g)}[0^{(g)}]$) could be defined for every genus\footnote{In \cite{CDG}, where it was considered the case $g=3$,
this form is called $G[0]$.} $g\geq 3$ ($g\geq 4$) considering three (four) dimensional isotropic subspace of
$\FF_2^{2g}$. In the same way, we can consider two dimensional isotropic subspaces of $\FF_2^{2g}$, for $g\geq 2$
and introduce another $O^+$-invariant, $G_2^{(g)}[0^{(g)}]$ (clearly
not linear independent from the others), that is a sum of suitable products of four theta constants at the fourth power.
This form will appear in the factorisation of some $O^+$-invariants.
%%%%%%%%%%%%%%%%%%%%%%%%%%%%%%%%%%%%%%%%%%%%%%%%%%%%%%%%%%%%%%%%%%%%%%%%%%%%%%%%%%%%%%%%%%%%%%%%%%%%%%%%%%%%%%%%%%%%%%

\section{Factorization of the $O^+$-invariants}\label{factorO}
\subsection{Genus one formulae} \label{1formula}
For the construction of the forms $\Xi^{(g)}_8[0^{(g)}]$ defining the chiral measure and to check that they have
the correct restriction on $\HH_1\times\HH_{g-1}$, it will be useful
to recall some identities between theta
constants at genus one. We will use the Dedekind function $\eta$ for which the classical formula
$\eta^3=\theta[{}_0^0]\theta[{}^0_1]\theta[{}^1_0]$
holds\footnote{Note that our definition of the Dedekind function
  differs from the classical ones, cf. \cite{RL}, for a factor $\frac
  12$. This explains the difference for a global factor $2^{4g}$
between our definition of the forms $\Xi_8^{(g)}[0^{(g)}]$ and the ones
in \cite{OPSY}.},
so $3\eta^{12}=\theta[{}^0_0]^{12}-\theta[{}^0_1]^{12}-\theta[{}^1_0]^{12}$. Also we consider the function
$f_{21}=2\theta[{}^0_0]^{12}+\theta[{}^0_1]^{12}+\theta[{}^1_0]^{12}$ as in \cite{CDG}. The two functions
$\eta^{12}$ and $f_{21}$ are a basis for the genus one $O^+$-anti-invariants \cite{DG} and so we can expand the anti-invariants
on this basis and the $O^+$-invariants on the basis $\theta[{}_0^0]^4\eta^{21}$, $\theta[{}_0^0]^4f_{21}$ and $F_{16}^{(1)}=\theta[{}_0^0]^{16}+\theta[{}_0^1]^{16}+\theta[{}_1^0]^{16} $. The proof of
these identities is straightforward using the Jacobi identity $\theta[{}_0^0]^4=\theta[{}^0_1]^4+\theta[{}^1_0]^4$:
\begin{align*}
&\theta^{12}[{}^0_0]=\frac{1}{3}f_{21}+\eta^{12}, \hskip 4.5cm \theta[{}^0_0]^4( \theta[{}^0_0]^8+\theta[{}^0_1]^8+\theta[{}^1_0]^8)
=\frac{2}{3}f_{21},\\
&\theta[{}^0_0]^{12}+\theta[{}^0_1]^{12}+\theta[{}^1_0]^{12}
=\frac{2}{3}f_{21}-\eta^{12}, \hskip 1.6cm \theta[{}^0_0]^4\theta[{}^0_1]^8+\theta[{}^0_0]^4\theta[{}^1_0]^8
=\frac{1}{3}f_{21}-\eta^{12},\\
&\theta[{}^0_0]^{16}+2\theta[{}^0_0]^{8}\theta[{}^0_1]^8+2\theta[{}^0_0]^{8}\theta[{}^1_0]^8=\theta[{}^0_0]^4(f_{21}-\eta^{12}), \\
&\frac{1}{2}\theta[{}^0_0]^{8}\theta[{}^0_1]^8+\frac{1}{2}\theta[{}^0_0]^{8}\theta[{}^1_0]^8-\theta[{}^0_0]^{4}\theta[{}^0_1]^{12}
-\theta[{}^0_0]^{4}\theta[{}^1_0]^{12}=\theta[{}^0_0]^4\left(\frac{3}{2}\eta^{12}-\frac{1}{6}f_{21}\right),\\
&\theta[{}^0_0]^{16}+3\theta[{}^0_0]^{8}\theta[{}^0_1]^8+3\theta[{}^0_0]^{8}\theta[{}^1_0]^8-2\theta[{}^0_0]^{4}\theta[{}^0_1]^{12}
-2\theta[{}^0_0]^{4}\theta[{}^1_0]^{12}=\theta[{}^0_0]^4\left(\frac{2}{3}f_{21}+2\eta^{12}\right),\\
&\theta[{}^0_1]^8\theta[{}^1_0]^8=\frac{1}{2}\theta[{}^0_1]^{16}+\frac{1}{2}\theta[{}^1_0]^{16}
+\frac{1}{2}\theta[{}^0_0]^8\theta[{}^0_1]^8+\frac{1}{2}\theta[{}^0_0]^8\theta[{}^1_0]^8-\theta[{}^0_0]^4\theta[{}^0_1]^{12}
-\theta[{}^0_0]^4\theta[{}^1_0]^{12} \\
&\phantom{\theta[{}^0_1]^8\theta[{}^1_0]^8}=\frac 12F_{16}^{(1)}+\theta[{}^0_0]^4(\eta^{12}-\frac 13f_{21}).
\end{align*}
%Note that the last one could be expand on the basis of $O^+$-invariant as $\theta[{}^0_1]^8\theta[{}^1_0]^8=1/2F_{16}^{(1)}+\theta[{}_0^0]^4\eta^{21}-1/3\theta[{}_0^0]^4f_{21}$, but the expression we gave (even if not on the basis vector) is more useful for what follows.

%%%%%%%%%%%%%%%%%%%%%%%%%%%%%%%%%%%%%%%%%%%%%%%%%%%%%%%%%%%%%%%%%%%%%%%%%%%%%%%%%%%%%%%%%%%%%%%%%%%%%%%%%%%%%%%%%%%
\subsection{The restrictions on $\HH_1\times\HH_{g-1}$}
Let us report here the factorisation of the six $O^+$-invariants found before for a reducible period matrix:
\begin{align*}
F^{(g)}_{1\,{|\Delta_{1,g-1}}}=&
\theta[{}_0^0]^4(\mbox{$\frac{1}{3}$}f_{21}+\eta^{12})F_1^{(g-1)},\\
F^{(g)}_{2\,{|\Delta_{1,g-1}}}
=&\theta[{}_0^0]^4(\mbox{$\frac{2}{3}$}f_{21}-\eta^{12})F^{(g-1)}_{2},\\
F^{(g)}_{3\,{|\Delta_{1,g-1}}}=&
\mbox{$\frac{2}{3}$}\theta[{}_0^0]^4f_{21}
F^{(g-1)}_{3},\\
F^{(g)}_{8\,{|\Delta_{1,g-1}}}=&
(\theta[{}^0_0]^{16}+\theta[{}^0_1]^{16}+\theta[{}^1_0]^{16}+2\theta[{}^0_0]^8\theta[{}^0_1]^8+2\theta[{}^0_0]^8\theta[{}^1_0]^8
+2\theta[{}^0_1]^8\theta[{}^1_0]^8)F_8^{(g-1)} \\
=&2F_{16}^{(1)}F_8^{(g-1)},
\\
F^{(g)}_{88\,{|\Delta_{1,g-1}}}
=&(\theta[{}^0_0]^{16}+\theta[{}^0_1]^{16}+\theta[{}^1_0]^{16}+2\theta[{}^0_0]^8\theta[{}^0_1]^8+2\theta[{}^0_0]^8\theta[{}^1_0]^8
-2\theta[{}^0_1]^8\theta[{}^1_0]^8)F^{(g-1)}_{88}\\
&+\theta[{}^0_1]^8\theta[{}^1_0]^8F_8^{(g-1)} \\
=&\theta[{}^0_1]^4f_{21}(\frac 43 F_{88}^{(g-1)}-\frac
13F_8^{(g-1)})+\theta[{}^0_1]^4\eta^{12}(-4F_{88}^{(g-1)}+F_8^{(g-1)})+\frac 12F_{16}^{(1)}F_8^{(g-1)}
,\\
F^{(g)}_{16\,{|\Delta_{1,g-1}}}=&
(\theta[{}^0_0]^{16}+\theta[{}^0_1]^{16}+\theta[{}^1_0]^{16}) F_{16}^{(g-1)}=F_{16}^{(1)}F_{16}^{(g-1)}.
\end{align*}
The factorisation of the forms $G_3^{(g)}[0^{(g)}]$ and $G_4^{(g)}[0^{(g)}]$ can be determined for any $g$ using
Theorem 15 of \cite{Grr} and we report the result for the cases $g=4,5$ in Section \ref{g4} and \ref{g5} respectively.

%%%%%%%%%%%%%%%%%%%%%%%%%%%%%%%%%%%%%%%%%%%%%%%%%%%%%%%%%%%%%%%%%%%%%%%%%%%%%%%%%%%%%%%%%%%%%%%%%%%%%%%%%%%%%%%%%%%%%%%%%%%%%%%%%
\section{Genus three case}\label{g3}
In Section \ref{f88} we found a basis for the five dimensional space of $O^+$-invariants satisfying the
transformation constraints. Let us search a linear combination satisfying the factorisation constraints. We will follow
the strategy of \cite{CDG}: write the more general vector in this space,
\begin{equation} \label{g3comblin}
\Xi^{(3)}_8[0^{(3)}]=a_1F^{(3)}_1+a_2F^{(3)}_2+a_3F^{(3)}_3+a_4F^{(3)}_8+a_5F^{(3)}_{88},
\end{equation}
and then impose for it to factorise as the
product of the genus one form
$\Xi^{(1)}_8[0^{(1)}]=\theta[{}^0_0]^4\eta^{12}$ and the form
$\Xi^{(2)}_8[0^{(2)}]=\frac 23 F_1^{(2)}+\frac 13F_2^{(2)}-\frac
12F_3^{(2)}$ at genus two, see \cite{CDG}, Section 3.4. In this way we obtain a linear equation
in the five coefficients $a_i$.

%%%%%%%%%%%%%%%%%%%%%%%%%%%%%%%%%%%%%%%%%%%%%%%%%%%%%%%%%%%%%%%%%%%%%%%%%%%%%%%%%%%%%%%%%%%%%%%%%%%%%%%%%%%%%%%%%%%%%%%%%%%%%%%%%%%
\subsection{The restriction of $\Xi^{(3)}_8[0^{(3)}]$ on $\HH_1\times\HH_2$}
The factorisation of the expression \eqref{g3comblin} for a reducible period matrix of the form
$\tau_{1,2}=\left(\begin{smallmatrix} \tau_1 & 0 \\ 0 & \tau_2\end{smallmatrix}\right)$ is:
\begin{align*}
&\left(a_1F^{(3)}_1+a_2F^{(3)}_2+a_3F^{(3)}_3+a_4F^{(3)}_8+a_5F^{(3)}_{88}\right)(\tau_{1,2})\\
&=\theta[{}^0_0]^4\left(a_1(\mbox{$\frac{1}{3}$}f_{21}+\eta^{12})F_1^{(2)}+a_2(\mbox{$\frac{2}{3}$}f_{21}-\eta^{12})F^{(2)}_{2}
+a_3\mbox{$\frac{2}{3}$}f_{21}F^{(2)}_{3}\right)
\\
&+a_42F_{16}^{(1)}F_8^{(2)}+a_5\left[\theta[{}^0_1]^4f_{21}(\frac 43 F_{88}^{(2)}-\frac
13F_8^{(2)})+\theta[{}^0_1]^4\eta^{12}(-4F_{88}^{(2)}+F_8^{(2)})%\right.
%\\&\left. 
+\frac 12F_{16}^{(1)}F_8^{(2)}\right]
\end{align*}
%This restriction takes the form
Necessary condition for the restriction to take the form:
\begin{align*}
&(\Xi^{(3)}_8[0^{(3)}])(\tau_{1,2})=\\
&\left(\theta[{}^0_0]^4\eta^{12}\right)(\tau_1)\Xi^{(2)}_8[0^{(2)}](\tau_2)\equiv \left(\theta[{}^0_0]^4\eta^{12}\right)(\tau_1)\left(\theta[0^{(2)}]^4
\Xi_6[0^{(2)}]\right)(\tau_2)
\end{align*}
%if the
is that the
terms proportional to $f_{21}$ and to $F_{16}^{(1)}$ disappear.
Here $\Xi_6[0^{(2)}]$ is the function found by D'Hoker and Phong in \cite{DP}.
First, let us impose the condition to get rid of the terms proportional to $f_{21}$.
This condition is satisfied if:
\bes
a_1\frac{1}{3}F_1^{(2)}+a_2\frac{2}{3}F_2^{(2)}+a_3\frac{2}{3}F_3^{(2)}
+a_5\left(\frac{4}{3}F_{88}^{(2)}-\frac{1}{3}F_8^{(2)}\right)=0.
\ees
This equation has a unique solution up to a scalar multiple:
\bes \label{solg3}
(a_1,a_2,a_3,a_5)=\lambda\left(\frac{16}{3},\frac{16}{3},-4,1\right),\qquad\qquad \lambda\in\CC.
\ees
To eliminate the terms proportional to $F_{16}^{(1)}$ the expression
$(2a_4+\frac{1}{2}a_5)F_8^{(2)}$ must vanish,
so, from the solution \ref{solg3}, we obtain
\bes
a_4=-\frac 14 a_5=-\lambda\frac 14.
\ees
Thus the expression for the factorised measure is:
\bes
\theta[{}_0^0]^4\eta^{12}\,\lambda\left[\frac{16}{3}F_1^{(3)}+\frac{16}{3}(-F_2^{(3)})+\left(F_{8}^{(3)}-4F_{88}^{(3)}\right)\right],
\ees
and it is of the form
$\Xi^{(3)}_8[0^{(3)}](\tau_{1,2})=(\theta[{}^0_0]^4\eta^{12})(\tau_1)(\theta[0^{(2)}]^4\Xi_6[0^{(2)}])$
if $\lambda=\frac{1}{16}$, as can be verified with a computer or using the classical theta formula.
This solution for the form $\Xi^{(3)}_8[0^{(3)}]$ is, up to a term proportional to $J^{(3)}$, the same found in
\cite{CDG}. Using this basis, the theta constants in the function $\Xi^{(3)}_8[0^{(3)}]$ appear with higher power than
using the one of \cite{CDG}: the four functions $F_1^{(3)}$, $F_3^{(3)}$, $F_8^{(3)}$ and $F_{88}^{(3)}$ are polynomials
in $\theta[\Delta^{(3)}]^8$ and they belong to $\sym^2({\bf 1}+{\bf 35}_b)$ and the $F_2^{(2)}$ is a polynomial in
$\theta[\Delta^{(3)}]^4$ and belongs to $\sym^4({\bf 15}_a)$.
The final expression for the form $\Xi^{(3)}_8[0^{(3)}]$ is:
\bes
\Xi^{(3)}_8[0^{(3)}]=\frac{1}{3}F^{(3)}_1+\frac{1}{3}F^{(3)}_2-\frac{1}{4}F^{(3)}_3-\frac{1}{64}F^{(3)}_8+\frac{1}{16}F^{(3)}_{88},
\ees
and it is completely equivalent to the one determined in \cite{CDG}:
\bes
\Xi^{(3)}_8[0^{(3)}]=\frac{1}{3}F^{(3)}_1+\frac{1}{3}F^{(3)}_2-\frac{1}{4}F^{(3)}_3-G_3^{(3)}[0^{(3)}].
\ees
In fact, they differ for a multiple of the form $J^{(3)}$ (that vanishes on the whole $\HH_3$), precisely the new expression is equal the old one $-\frac{5}{448}J^{(3)}$, as it can be computed using the results of Table \ref{tab:sum3}. This procedure will be explained in detail for the genus four case in the next Section.
So we can use this two expressions to show that the form $G_3^{(3)}[0^{(3)}]$ is polynomial in $\theta[\Delta^{(3)}]^8$:
\bes
G_3^{(3)}[0^{(3)}]=\frac{1}{64}F_8^{(3)}-\frac{1}{16}F_{88}^{(3)}-\frac{5}{448}(8F_{16}^{(3)}-F_8^{(3)}).
\ees
We include also the form $J^{(3)}$ because it is zero as a function of $\tau\in\HH_3$ and using a computer to perform the computations one has to add explicitly this fact.

%%%%%%%%%%%%%%%%%%%%%%%%%%%%%%%%%%%%%%%%%%%%%%%%%%%%%%%%%%%%%%%%%%%%%%%%%%%%%%%%%%%%%%%%%%%%%%%%%%%%%%%%%%%%%%%%%%%%%%%%%%%%%%%%%%%%%%%%%%%%%%
\section{Genus four case}\label{g4}
For the genus four case we repeat the method used for the genus three in the previous Section. In Section \ref{oinvg4}
we found seven linear independent $O^+$-invariants that form a basis. We can now search a linear combination
that also satisfies the factorisation constraints:
\begin{equation}\label{g4comblin}
\Xi^{(4)}_8[0^{(4)}]=a_1F^{(4)}_1+a_2F^{(4)}_2+a_3F^{(4)}_3+a_4F^{(4)}_8+a_5F^{(4)}_{88}+a_6F_{16}^{(4)}+a_7G_3^{(4)}[0^{(4)}],
\end{equation}
where $G_3^{(4)}[0^{(4)}]$ is the function defined in \ref{oinvg4}, or in the notations of Grushevsky $P_{3,2}^{(4)}$. In this case we also use the $F_{16}^{(4)}$ because in $g=4$ it is
independent from $F_8^{(4)}$, i.e. the expression $J^{(4)}$ is not identically zero on the whole $\HH_4$, but just on the Jacobi locus.
%%%%%%%%%%%%%%%%%%%%%%%%%%%%%%%%%%%%%%%%%%%%%%%%%%%%%%%%%%%%%%%%%%%%%%%%%%%%%%%%%%%%%%%%%%%%%%%%%%%%%%%%%%%%%%%%%%%%%%%%%%%%%%%%%%%%%%%%%%%%%%%%
\subsection{The restriction of $\Xi^{(4)}_8[0^{(4)}]$ on $\HH_1\times\HH_3$}
The restriction  on $\HH_1\times\HH_3$ of the function $G_3^{(4)}[0^{(4)}]$ was found in \cite{CDG2}:
\begin{eqnarray*}
G_3^{(4)}[0^{(4)}](\tau_{1,3})&=& \theta[{}_0^0]^4(\tau_1)\left[\frac{1}{3}f_{21}(\tau_1)\left(G_2^{(3)}[0^{(3)}]+8G_3^{(3)}[0^{(3)}]\right)(\tau_3)\right. \\
& &\left.-\eta^{12}(\tau_1)\left(G_2^{(3)}[0^{(3)}]+6G_3^{(3)}[0^{(3)}]\right)(\tau_3)\right],
\end{eqnarray*}
this follows also from Theorem 15 of \cite{Grr}.
The modular forms $G_3^{(3)}[0^{(3)}]$ and $G_2^{(3)}[0^{(3)}]$ are defined in \cite{CDG} and \cite{CDG2} respectively.
Thus the factorisation of the expression \eqref{g4comblin} for a reducible period matrix of the form
$\tau_{1,3}=\left(\begin{smallmatrix} \tau_1 & 0 \\ 0 & \tau_3\end{smallmatrix}\right)$ is:
\begin{align*}
&\left(a_1F^{(4)}_1+a_2F^{(4)}_2+a_3F^{(4)}_3+a_4F^{(4)}_8+a_5F^{(4)}_{88}+a_6F_{16}^{(4)}+a_7G_3^{(4)}[0^{(4)}]\right)(\tau_{1,3})\\
&=\theta[{}^0_0]^4\left[a_1(\mbox{$\frac{1}{3}$}f_{21}+\eta^{12})F_1^{(3)}+a_2(\mbox{$\frac{2}{3}$}f_{21}-\eta^{12})F^{(3)}_{2}
+a_3\mbox{$\frac{2}{3}$}f_{21}F^{(3)}_{3}\right]+a_42F_{16}^{(1)}F_8^{(3)}\\
&+a_5\left[\theta[{}^0_1]^4f_{21}(\frac 43 F_{88}^{(3)}-\frac
13F_8^{(3)})+\theta[{}^0_1]^4\eta^{12}(-4F_{88}^{(3)}+F_8^{(3)})%\right.
%\\&\left. 
+\frac 12F_{16}^{(1)}F_8^{(3)}\right]\\
&+a_6F_{16}^{(1)}F_{16}^{(3)}+
%&a_4\left[\theta[{}^0_0]^{16}+\theta[{}^0_1]^{16}+\theta[{}^1_0]^{16}+2\theta[{}^0_0]^8\theta[{}^0_1]^8+2\theta[{}^0_0]^8\theta[{}^1_0]^8
%+2\theta[{}^0_1]^8\theta[{}^1_0]^8\right]F_8^{(3)}+\\
%&a_5\left[(\theta[{}^0_0]^{16}+\theta[{}^0_1]^{16}+\theta[{}^1_0]^{16}+2\theta[{}^0_0]^8\theta[{}^0_1]^8+2\theta[{}^0_0]^8\theta[{}^1_0]^8
%-2\theta[{}^0_1]^8\theta[{}^1_0]^8)F^{(3)}_{88}+\theta[{}^0_1]^8\theta[{}^1_0]^8F_8^{(3)}\right]+\\
%&a_6\left[\theta[{}^0_0]^{16}+\theta[{}^0_1]^{16}+\theta[{}^1_0]^{16}\right]F_{16}^{(3)}+\\
a_7\left[\theta[{}^0_0]^{4}
  f_{21}(\frac{1}{3}G_2^{(3)}[0^{(3)}]+\frac{8}{3}G_3^{(3)}[0^{(3)}])\right.\\
&\left.+\theta[{}^0_0]^{4}\eta^{12}(-G_2^{(3)}[0^{(3)}]-6G_3^{(3)}[0^{(3)}])\right].
\end{align*}
The terms proportional to $f_{21}$ disappear if:
\begin{align*}
&a_1\frac{1}{3}F_1^{(3)}+a_2\frac{2}{3}F_2^{(3)}+a_3\frac{2}{3}F_3^{(3)}
+a_5\left(\frac{4}{3}F_{88}^{(3)}-\frac{1}{3}F_8^{(3)}\right)\\
&+a_7\frac{1}{3}(G_2^{(3)}[0^{(3)}]+8G_3^{(3)}[0^{(3)}])=0.
\end{align*}
This equation has a unique solution up to a scalar multiple:
\bes
(a_1,a_2,a_3,a_5,a_7)=\lambda\left(-\frac{56}{5},-\frac{112}{5},\frac{42}{5},-\frac{21}{5},\frac{168}{5}\right),\qquad\qquad \lambda\in\CC.
\ees
The term proportional to $F_{16}^{(1)}$ vanishes if:
\bes
a_42F_8^{(3)}+a_5\frac 12 F_8^{(3)}+a_6F_{16}^{(3)}=0.
\ees
This equation has infinitely many solutions. Due to the vanishing of
$J^{(3)}$ on the whole Siegel upper half space we can rewrite the
previous equation as $(2a_4+\frac 12 a_5 +\frac 18 a_6)F_8=0$, which has
solution $a_4=-\frac{a_5}{4}-\frac{a_6}{16}$ with $a_6\in \CC$. For
any choice of $a_6$ an additive term proportional to $J^{(4)}$
appears in the expression of $\Xi_8^{(4)}[0^{(4)}]$ and precisely it
is $\frac{a_6}{16}(16F_{16}^{(4)}-F_8^{(4)})$. In this sense the form
$\Xi_8^{(4)}[0^{(4)}]$ is unique up to a term proportional to
$J^{(4)}$, as proved in \cite{DG}. Thus, we can choose $a_6=0$ and
$a_4=-\frac 14 a_5$. The request for the restriction to be of the
form $\Xi^{(4)}_8[0^{(4)}](\tau_{1,3})=\left(\theta[{}^0_0]^4\eta^{12}\right)(\tau_1)\Xi^{(3)}_8[0^{(3)}](\tau_3)$
fixes the value of $\lambda=-\frac {5}{336}$. This follow from the condition:
\begin{align*}
\theta[{}_0^0]^4\eta^{12}\,\lambda&\left[-\frac{56}{5}F_1^{(3)}-\frac{112}{5}(-F_2^{(3)})+\frac{21}{5}\left(F_{8}^{(3)}-4F_{88}^{(3)}\right)\right.\\
&\phantom{[\,}\left.+\frac{168}{5}\left[-\left(G_2^{(3)}[0^{(3)}]+6G_3^{(3)}[0^{(3)}]\right)\right]\right]=\theta[{}_0^0]^4\eta^{12}\Xi_8^{(3)}[0^{(3)}],
\end{align*}
and, using again the fact that $J^{(3)}$ identically vanishes, we
obtain $\lambda=-\frac {5}{336}$.

The above discussion shows that the form $\Xi^{(4)}_8[0^{(4)}]$ is:
\begin{equation} \label{xi8g4n}
\Xi^{(4)}_8[0^{(4)}]=\frac{1}{6}F^{(4)}_1+\frac{1}{3}F^{(4)}_2-\frac{1}{8}F^{(4)}_3+\frac{1}{64}F^{(4)}_8-\frac{1}{16}F^{(4)}_{88}-\frac
12G_3^{(4)}[0^{(4)}], 
\end{equation}
which, for the uniqueness (up to a multiple of $J^{(4)}$) of the form $\Xi_8^{(4)}[0^{(4)}]$, is equivalent to the one found in \cite{CDG2}:
\begin{equation}\label{xi8g4o}
\Xi^{(4)}_8[0^{(4)}]=\frac{1}{6}F^{(4)}_1+\frac{1}{3}F^{(4)}_2-\frac{1}{8}F^{(4)}_3-\frac{1}{2}G_3^{(4)}[0^{(4)}]+4G_4^{(4)}[0^{(4)}].
\end{equation}
This two expressions must be equated and they could differ just for a multiple of the Schottky relation $J^{(4)}$. Calling the form $\Xi_8^{(4)}[0^{(4)}]$ \eqref{xi8g4n} found in this work $\Xi_8^{(4)}[0^{(4)}]_{DP}$ and the \eqref{xi8g4o} of \cite{CDG2} $\Xi_8^{(4)}[0^{(4)}]_{CDG}$ we can write:
\bes
\Xi_8^{(4)}[0^{(4)}]_{DP}=\Xi_8^{(4)}[0^{(4)}]_{CDG}+aJ^{(4)}.
\ees
Summing over the 136 even theta characteristics and using the results of Table \ref{tab:sum4} we find:
\bes
\frac{45}{56}J^{(4)}=\frac{12}{7}J^{(4)}+136aJ^{(4)},
\ees
from which it follows that $a=-3/448$. This shows that the modular form $G_4^{(4)}[0^{(4)}]$ is, actually, polynomial in
$\theta[\Delta^{(4)}]^8$:
\begin{align}
G_4^{(4)}[0^{(4)}]&=\frac{1}{256}F^{(4)}_8-\frac{1}{64}F_{88}^{(4)}+\frac{3}{1792}J^{(4)}\\ \nonumber
&=\frac{1}{448}F_8^{(4)}-\frac{1}{64}F_{88}^{(4)}+\frac{3}{112}F_{16}^{(4)}.
\end{align}
\subsubsection{Remark}\label{smresult}
Recently, Oura has proved, as a consequence of the results in
\cite{Ng}, that the space of modular forms with respect to the
subgroup $\Gamma_4(1,2)$ of weight 8, quadratic in the theta
constants, has dimension no bigger than 7, $\dim M_8 ^{\theta^2}
(\Gamma_4(2))^{O^+}\leq ~ 7$.
%$\dim[\Gamma_4(1,2),8]^{\theta^2}\leq ~7$.
The computations in the previous Section show that this dimension is precisely seven:
\bes
\dim M_8 ^{\theta^2}
(\Gamma_4(2))^{O^+}=7.
%\dim[\Gamma_4(1,2),8]^{\theta^2}= 7.
\ees
Moreover, in \cite{OPSY} it is proved that the space of cusp forms $[\Gamma_4(1,2),8]_0$, in which $\Xi_8^{(4)}[0^{(4)}]$ lies, has dimension two. From this, it follows the uniqueness (up to a multiple of $J^{(4)}$) of the form $\Xi_8^{(4)}[0^{(4)}]$ (as explained in \cite{OPSY} and \cite{GS}) and not just in a weakened form, i.e. assuming polynomiality in the theta constants, as in \cite{DG}.

\section{Genus five case}\label{g5}
In Section \ref{oinvg5} we found eight linear independent $O^+$-invariants that form a basis. Their general linear combination is:
\begin{align} \label{g5comblin}
\Xi^{(5)}_8[0^{(5)}]&=a_1F^{(5)}_1+a_2F^{(5)}_2+a_3F^{(5)}_3+a_4F^{(5)}_8+a_5F^{(5)}_{88}\\
&+a_6F_{16}^{(5)}+a_7G_3^{(5)}[0^{(5)}]+a_8G_4^{(5)}[0^{(5)}]. \nonumber
\end{align}
We will search for eight coefficients $a_i$ such that this expression satisfies the right factorisation.
\subsection{The restriction of $G_3^{(5)}[0^{(5)}]$ and $G_4^{(5)}[0^{(5)}]$ on $\HH_1\times\HH_4$}
These restrictions follow quite directly from the Theorem of Grushevsky, identifying
$G_4^{(5)}[0^{(5)}]$ with $P_{4,1}^{(5)}$ and $G_3^{(5)}[0^{(5)}]$ with $P_{3,2}^{(5)}$.
For the function $G_3^{(5)}[0^{(5)}]$ we get:
\begin{align*}
P_{3,2}^{(5)}(\tau_{1,4})&=P_{0,16}^{(1)}(\tau_1)P_{3,2}^{(4)}(\tau_4)+P_{1,8}^{(1)}(\tau_1)P_{2,4}^{(4)}(\tau_4)+7P_{1,8}^{(1)}(\tau_1)P_{3,2}^{(4)}(\tau_4) \\
&=\theta[{}^0_0]^4(\frac{8}{3}f_{21}-6\eta^{12})(\tau_1)G_3^{(4)}[0^{(4)}](\tau_4)+\theta[{}^0_0]^4(\frac{1}{3}f_{21}-\eta^{12})(\tau_1)G_2^{(4)}[0^{(4)}](\tau_4),
\end{align*}
where we used $P_{0,16}^{(1)}=\theta[0^{(1)}]^4\left(\frac{1}{3}f_{21}+\eta^{12}\right)$,
$P_{1,8}^{(1)}=\theta[0^{(1)}]^4\left(\frac{1}{3}f_{21}-\eta^{12}\right)$ and $P_{2,4}^{(4)}=G_2^{(4)}[0^{(4)}]$.
For $G_4^{(5)}[0^{(5)}]$ we get:
\begin{align*}
P_{4,1}^{(5)}(\tau_{1,4})&=P_{0,16}^{(1)}(\tau_1)P_{4,1}^{(4)}(\tau_4)+P_{1,8}^{(1)}(\tau_1)P_{3,2}^{(4)}(\tau_4)+15P_{1,8}^{(1)}(\tau_1)P_{4,1}^{(4)}(\tau_4) \\
&=\theta[{}^0_0]^4(\frac{16}{3}f_{21}-14\eta^{12})(\tau_1)G_4^{(4)}[0^{(4)}](\tau_4)+\theta[{}^0_0]^4(\frac{1}{3}f_{21}-\eta^{12})(\tau_1)G_3^{(4)}[0^{(4)}](\tau_4),
\end{align*}
where, as before, $P_{0,16}^{(1)}=\theta[0^{(1)}]^4\left(\frac{1}{3}f_{21}+\eta^{12}\right)$ and
$P_{1,8}^{(1)}=\theta[0^{(1)}]^4\left(\frac{1}{3}f_{21}-\eta^{12}\right)$.
These restrictions could be also determined using the method of isotropic subspaces, as in \cite{CDG}, or by
direct computation  using a computer. In the next Section it will be useful to use also
$G_2^{(4)}[0^{(4)}]=\left(2F_1^{(4)}+16F_2^{(4)}-3F_3^{(4)}\right)/6$ (instead, at genus three we have
$G_2^{(3)}[0^{(3)}]=\left(2F_1^{(3)}+8F_2^{(3)}-3F_3^{(3)}\right)/6$).

\subsection{The restriction of $\Xi^{(5)}_8[0^{(5)}]$ on
  $\HH_1\times\HH_4$} \label{resg5-14}
Using the results of the previous Section and of Section \ref{factorO} the factorisation of the expression
\eqref{g5comblin} for a reducible period matrix of the form
$\tau_{1,4}=\left(\begin{smallmatrix} \tau_1 & 0 \\ 0 & \tau_4\end{smallmatrix}\right)$ is:
\begin{align*}
&\left(a_1F^{(5)}_1+a_2F^{(5)}_2+a_3F^{(5)}_3+a_4F^{(5)}_8+a_5F^{(5)}_{88} \right. \\&
\qquad
\qquad\left. +a_6F_{16}^{(5)}+a_7G_3^{(5)}[0^{(5)}]+a_8G_4^{(5)}[0^{(5)}]\right)(\tau_{1,4})=\\&
\left[a_1 \theta[{}^0_0]^4 (\mbox{$\frac{1}{3}$}f_{21}+\eta^{12})F_1^{(4)}+a_2 \theta[{}^0_0]^4 (\mbox{$\frac{2}{3}$}f_{21}-\eta^{12})F^{(4)}_{2}+a_3\mbox{$\frac{2}{3}$}\theta[{}^0_0]^4 f_{21}F^{(4)}_{3}\right]+
\\ & a_42F_{16}^{(1)}F_8^{(4)}+a_5\left[\theta[{}^0_1]^4f_{21}(\frac 43 F_{88}^{(4)}-\frac
13F_8^{(4)})+\theta[{}^0_1]^4\eta^{12}(-4F_{88}^{(4)}+F_8^{(4)})\right.
\\&
\left. +\frac 12F_{16}^{(1)}F_8^{(4)}\right]+a_6F_{16}^{(1)}F_{16}^{(4)}+\\&
a_7\theta[{}^0_0]^4\left[\frac{1}{3}f_{21}(G_2^{(4)}[0^{(4)}]+8G_3^{(4)}[0^{(4)}])+\eta^{12}(-G_2^{(4)}[0^{(4)}]-6G_3^{(4)}[0^{(4)}])\right]+\\&
a_8\theta[{}^0_0]^4\left[f_{21}\left(\frac{16}{3}G_4^{(4)}[0^{(4)}]+\frac
    13
    G_3^{(4)}[0^{(4)}]\right)+\eta^{12}\left(-14G_4^{(4)}[0^{(4)}]-G_3^{(4)}[0^{(4)}]\right)\right]. \\&
\end{align*}
From the relations of paragraph \ref{1formula} we obtain the condition for the vanishing of the terms proportional to
$f_{21}$:
\begin{align*}
a_1\frac{1}{3}F_1^{(4)}+a_2\frac{2}{3}F_2^{(4)}+a_3\frac{2}{3}F_3^{(4)}+a_5\left(\frac{4}{3}F_{88}^{(4)}-\frac{1}{3}F_8^{(4)}\right)&+\\
+a_7\frac{1}{3}(G_2^{(4)}[0^{(4)}]+8G_3^{(4)}[0^{(4)}])+a_8\frac{1}{3}(G_3^{(4)}[0^{(4)}]+16G_4^{(4)}[0^{(4)}])&=0.
\end{align*}
Again, this equation, using the fact that $J^{(4)}$ vanishes on the
Jacobi locus (actually, one solves the equation modulo the Schottky
relation $J^{(4)}$), has an unique solution, up to a scalar multiple:
\bes
(a_1,a_2,a_3,a_5,a_7,a_8)=\lambda\left(-\frac{14}{3},-\frac{56}{3},\frac{7}{2},-7,14,-112\right),\quad\qquad \lambda\in\CC.
\ees
The term proportional to $F_{16}^{(1)}$ vanishes if:
\bes
a_42F_8^{(4)}+a_5\frac 12 F_8^{(4)}+a_6F_{16}^{(4)}=0.
\ees
As for $g=4$ this equation has infinitely many
solutions and using again the fact that $J^{(4)}$ vanishes on the Jacobi locus we
obtain $(2a_4+~\frac{1}{2}a_5+~\frac{1}{16}a_6)F^{(4)}_8=0$, which has
solution $a_4=-\frac{a_5}{4}-\frac{a_6}{32}$, with $a_6\in\CC$. For any choice of the coefficient $a_6$ the additive term
$\frac{a_6}{32}(32F_{16}^{(5)}-F_8^{(5)})$ appears in
$\Xi_8^{(5)}[0^{(5)}]$. However, in genus five this term vanishes just
on the locus of trigonal curves and not on the whole Jacobi
locus. This shows how the uniqueness of the form
$\Xi_8^{(5)}[0^{(5)}]$ can not be longer assured by the three
constraints of Section \ref{strategy} on $J_5$, as also pointed out
in \cite{OPSY}. Thus, we can choose $a_6=0$ and $a_4=-\frac 14 a_5$.
The request for the restriction to be of the form
$\Xi^{(5)}_8[0^{(5)}](\tau_{1,4})=(\theta[{}_0^0]^4\eta^{12})(\tau_1)\Xi^{(4)}_8[0^{(4)}](\tau_4)$ means that:
\begin{align*}
&(\theta[{}_0^0]^4\eta^{12})(\tau_1)\lambda\left[a_1F_1^{(4)}-a_2F_2^{(4)}+a_5\left(F^{(4)}_8-4F^{(4)}_{88}\right)\right.\\
&\left. \phantom{(\theta[{}_0^0]^4\eta^{12})(\tau_1)\lambda}+a_7(-G_2^{(4)}[0^{(4)}]-6G_3^{(4)}[0^{(4)}])+a_8(-14G_4^{(4)}[0^{(4)}]-G_3^{(4)}[0^{(4)}])\right]
\\
&=(\theta[{}_0^0]^4\eta^{12})(\tau_1)\Xi^{(4)}_8[0^{(4)}],
%\ees
\end{align*}
where $\Xi^{(4)}_8[0^{(4)}]$ is
the function found in Section~ \ref{g4}, and this should fix the
constant $\lambda$. 
Therefore, we impose:
\begin{align*}
\theta[{}_0^0]^4\eta^{12}\,\lambda&\left[-\frac{14}{3}F_1^{(4)} -
  \frac{56}{3}(-F_2^{(4)}) - 7(F_8^{(4)} - 4F_{88}^{(4)}) +14 (-G_2^{(4)}[0^{(4)}]-6G_3^{(4)}[0^{(4)}])
\right.\\
&\left.-112(-14G_4^{(4)} - G_3^{(4)})\right]
=\theta[{}_0^0]^4\eta^{12}\left(\Xi_8^{(4)}[0^{(4)}]+\Lambda J^{(4)}\right),
\end{align*}
this equation has solution $\lambda=-\frac{1}{56}$ and
$\Lambda=-\frac{3}{64}$. Actually, using $\lambda=-\frac{1}{56}$ and summing over
all the even characteristics one finds that the expression in the
square brackets on the left and the $\Xi_8^{(4)}[0^{(4)}]$ on the right sides
of the previous equation differ by $-\frac{3}{64}J^{(4)}$.
Thus, in genus five the function $\Xi_8^{(5)}[0^{(5)}]$ satisfying
the three constraints on the Jacobi locus is:
\be \label{g5pf}
\Xi_8^{(5)}[0^{(5)}]=\frac{1}{12}F_1^{(5)}+\frac 13F_2^{(5)}-\frac
{1}{16}F_3^{(5)}-\frac{1}{32}F_8^{(5)}+\frac 18F_{88}^{(5)}-\frac 14 G_3^{(5)}[0^{(5)}]+2G_4^{(5)}[0^{(5)}].
\ee
Note that also for the solution found in \cite{OPSY} the correct
restriction holds if one restrict to $J_4$.

\subsection{The constraint on $\HH_2\times\HH_3$}
Now we consider the restriction of the function $\Xi^{(5)}_8[0^{(5)}]$
to $\HH_2\times\HH_3$ and this, to satisfy the factorization
constraint of Section \ref{strategy}, must be equal to the product
$\Xi^{(2)}_8[0^{(2)}]\Xi^{(3)}_8[0^{(3)}]$ i.e. the genus two
times the genus three measure.

In order to obtain the restriction of $\Xi^{(5)}_8[0^{(5)}]$ we need
the restriction of the eight basis functions. We have:
\begin{align*}
F^{(5)}_{1\,{|\Delta_{2,3}}}=&F_1^{(2)}F_1^{(3)},\\
F^{(5)}_{2\,{|\Delta_{2,3}}}
=&F^{(2)}_{2}F^{(3)}_{2},\\
F^{(5)}_{3\,{|\Delta_{2,3}}}=&F^{(2)}_{3}F^{(3)}_{3},\\
F^{(5)}_{8\,{|\Delta_{2,3}}}=&F_8^{(2)}F_8^{(3)},
\\
F^{(5)}_{16\,{|\Delta_{2,3}}}=&F_{16}^{(2)}F_{16}^{(3)},\\
F^{(5)}_{88\,{|\Delta_{2,3}}}
=&F_1^{(2)}(\frac{16}{3}F_{88}^{(3)}-\frac{4}{3}F_8^{(3)})+F_2^{(2)}(\frac{32}{3}F_{88}^{(3)}
-\frac 83 F_8^{(3)})\\
+&F_3^{(2)}(-8F_{88}^{(3)} +2F_8^{(3)})+F_{16}^{(2)}F_8^{(3)},
\\
G^{(5)}_{3\,{|\Delta_{2,3}}}=&G_0^{(2)}G_3^{(3)}+7
G_1^{(2)}G_3^{(3)}+G_1^{(2)}G_2^{(3)}+42 G_2^{(2)}G_3^{(3)}+9
G_2^{(2)}G_2^{(3)}+G_2^{(2)}G_1^{(3)}\\
=&F_1^{(2)}(\frac 13F_1^{(3)}+\frac 83 F_2^{(3)}-\frac 23
F_3^{(3)}+\frac 18 F_8^{(3)}-\frac 12 F_{88}^{(3)})\\
+&F_2^{(2)}(\frac 43
F_1^{(3)}+8F_2^{(3)}-\frac 73 F_3^{(3)}+\frac{7}{16}F_8^{(3)}-\frac 74
F_{88}^{(3)})\\
+&F_3^{(2)}(-\frac 23
F_1^{(3)}-\frac{14}{3}F_2^{(3)}+\frac 54
F_3^{(3)}-\frac{7}{32}F_8^{(3)}+\frac 78F_{88}^{(3)}),\\
G^{(5)}_{4\,{|\Delta_{2,3}}}=&G_1^{(2)}G_3^{(3)}+21G_2^{(2)}G_3^{(3)}+G_2^{(2)}G_2^{(3)}\\
=&F_1^{(2)}(\frac 19F_1^{(3)}+\frac 49 F_2^{(3)}-\frac 16
F_3^{(3)}+\frac {3}{32} F_8^{(3)}-\frac 38 F_{88}^{(3)})\\
+&F_2^{(2)}(\frac 29
F_1^{(3)}+\frac 89F_2^{(3)}-\frac 13 F_3^{(3)}+\frac{7}{32}F_8^{(3)}-\frac 78
F_{88}^{(3)})\\
+&F_3^{(2)}(-\frac 16
F_1^{(3)}-\frac{2}{3}F_2^{(3)}+\frac 14
F_3^{(3)}-\frac{19}{128}F_8^{(3)}+\frac {19}{32}F_{88}^{(3)}).
\end{align*}
The first five relations follow quite easly from the definitions and
the classical theta formula. The sixth is longer to prove in
the same manner and it can be obtained using software like
Mathematica. 
The last two follow from Theorem 15 of \cite{Grr}. In \cite{CDGd},
using the relations between the lattice theta series and the classical
theta constants, another proof of the first six restrictions and of
the one of $G_4^{(5)}[0^{(5)}]$ will be given.

We can now obtain the restriction of the form $\Xi_8^{(5)}[0^{(5)}]$:
\begin{align*}
\Xi_8^{(5)}[0^{(5)}](\tau_{2,3})&=\left(\frac 23 F_1^{(2)}+\frac 13
F_2^{(2)}-\frac 12F_3^{(2)}\right)\\
&\phantom{=}\cdot\left(\frac 13 F_1^{(3)} +\frac 13
F_2^{(3)}-\frac 14
F_3^{(3)}-\frac{1}{64}F_8^{(3)}+\frac{1}{16}F_{88}^{(3)}\right) \\
&=\Xi_8^{(2)}[0^{(2)}](\tau_2)\Xi_8^{(3)}[0^{(3)}](\tau_3).
\end{align*}
Therefore the modular form $\Xi_8^{(5)}[0^{(5)}]$ on $\Gamma_5(1,2)$
of weight 8,
defined in \ref{resg5-14}, satisfies all the factorization
constraints in genus five.

\subsubsection{Remark}
It is interesting to investigate the possibility to apply a similar procedure
to the genus six case. However, in this case we have no indication
about the dimension of the space of $O^+$-invariant modular forms, not
even if one
restricts oneself to the polynomial in the theta constants part of the ring of modular forms. Tentatively
one can try to built some invariants employing the action of the generators of the
symplectic group, as we did in Section \ref{f88} to define the
function $F_{88}^{(3)}$, and search a linear combination among them
with the correct restriction. These topics will be considered for a
future work.

%In fact, these two expressions are
%not proportional: there are no value for $\lambda$ that makes true this equality.
%This negative result proves the Theorem \ref{theo1} stated in Section \ref{intro} because this computation shows that the form
%$\Xi_8^{(5)}[0^{(5)}]$ cannot be written as a polynomial in the theta constants, but the presence of a term with a square root, like in \cite{Grr}, is inevitable and Salvati Manni in \cite{SM} showed that this term is well defined on the moduli space of curves.

\subsection{On the dimensions of certain space of modular forms}
In Sections \ref{g4} and \ref{g5} we considered the space of the modular forms of weight 8 with respect the group
$\Gamma_g(1,2)$. In particular we focused on the modular forms polynomial in theta constants. In order to find the
forms $\Xi_8^{(g)}[0^{(g)}]$ that factorise in the right way we searched for a basis for these spaces and this allowed us
to find the dimensions of the spaces. We summarise these results in
the following (cf. Remark \ref{eightfuncts}):
\begin{proposition}
For the space 
$M_8 ^{\theta^2} (\Gamma_4(2))^{O^+}$, $M_8 ^{\theta^2} (\Gamma_5(2))^{O^+}$ and $M_8 ^{\theta} (\Gamma_5(2))^{O^+}$
%$[\Gamma_4(1,2),8]^{\theta^2}$, $[\Gamma_5(1,2),8]^{\theta^2}$ and $[\Gamma_5(1,2),8]^{\theta}$
the following equalities hold:
\begin{align*}
&\dim M_8 ^{\theta^2} (\Gamma_4(2))^{O^+}=7, \\
&\dim M_8 ^{\theta^2} (\Gamma_5(2))^{O^+}=7, \\
&\dim M_8 ^{\theta} (\Gamma_5(2))^{O^+}=8.
\end{align*}
\end{proposition}

\section{The vanishing of the cosmological constant}
In this Section we reinterpret the vanishing of the cosmological constant on the light of the group representation
theory. In Section \ref{strategy} we pointed out that the $O^+$-invariants belong to the ${\bf 1}$ and
$\sigma_\theta$ representations. 
%So, the space of $O^+$-invariants decomposes as a direct sum as follows:
%\bes
%M_8(\Gamma_g(2))^{O^+}=n_1{\bf 1} \oplus n_{\sigma_\theta}\sigma_\theta.
%\ees
For the case $g\leq 5$ we know that the only $\modular(2g)$ invariants are $F_{16}^{(g)}$ and $F_8^{(g)}$ (they are
not independent for $g=3$) and they form a basis for the ${\bf 1}$
part of the space of the $O^+$-invarints. Let
$\{e_{\sigma_i}\}_{i=1,\cdots,n_{\sigma_\theta}}$ be the basis for the
$\sigma_\theta$ part.
Then, an $O^+$-invariant decomposes in two parts: the first one lying in the representation ${\bf 1}$ and the second one
in  the $\sigma_\theta$. Thus, if $f[0^{(g)}]\in M_8(\Gamma_g(2))^{O^+}$, we can write
$f[0^{(g)}]=aF_8^{(g)}+bF_{16}^{(g)}+\sum_i^{n_{\sigma_\theta}}c_ie_{\sigma_i}$, for $g\leq 5$. Acting on
these functions with all the generators of the group $\modular(2g)$ and summing up the result at each step we obtain a
$\modular(2g)$-invariant. We know that the unique $\modular(2g)$-invariants are the two functions $F_8^{(g)}$ and
$F_{16}^{(g)}$ so that the $\sigma_\theta$ representation part gives no contribution to the sum. Therefore, if
the function $f[0^{(g)}]$ contains a non trivial part proportional to $F_8^{(g)}$ or $F_{16}^{(g)}$, the result of the
sum will be non zero.

The cosmological constant is the sum of the functions $\Xi_8^{(g)}[\Delta^{(g)}]$ over all the even characteristics.
This sum is a $\modular(2g)$-invariant and it must then be proportional to a combination of $F_8^{(g)}$ and
$F_{16}^{(g)}$. Thus the cosmological constant vanishes if this sum is zero.
We now verify this for the genus three, four and five cases.

\subsection{Genus three}
In Table \ref{tab:sum3} we report the sums of each term appearing in the form $\Xi_8^{(3)}[0^{(3)}]$. These show that
for the expression of the measure in the three bases (the one of \cite{CDG} (CDG), the one in this work (DP) and the
basis of \cite{Grr} (Gr)) for the space of $O^+$-invariants we always obtain the vanishing of the cosmological constant
(as expected) due to the vanishing of the form $J^{(3)}$.

\begin{table}[h!]
\begin{center}
\resizebox*{1\textwidth}{!}{
\begin{tabular}{ccccccccc}
\toprule
Function && Sum && CDG && DP && Gr \\
\midrule
$F_1^{(3)}$ && $F_{16}^{(3)}$ && $\frac{1}{3}$ && $\frac{1}{3}$ && $\frac{1}{8}$\\
$F_2^{(3)}$ &&  $8F_{16}^{(3)}$ && $\frac{1}{3}$ && $\frac{1}{3}$ && 0\\
$F_3^{(3)}$ &&  $F_8^{(3)}$ && $-\frac{1}{4}$ && $-\frac{1}{4}$ && 0\\
$F_8^{(3)}$ && $36F_8^{(3)}$ && 0 &&$-\frac{1}{64}$ && 0\\
$F_{88}^{(3)}$ && $8F_8^{(3)}-8F_{16}^{(3)}$ && 0 && $\frac{1}{16}$ && 0\\
$G_1^{(3)}[0^{(3)}]$ && $F_{8}^{(3)}-F_{16}^{(3)}$ && 0 && 0 && $-\frac{1}{8}$\\
$G_2^{(3)}[0^{(3)}]$ && $11F_{16}^{(3)}-\frac{1}{2}F_{8}^{(3)}$ && 0 && 0 &&$\frac{1}{4}$\\
$G_3^{(3)}[0^{(3)}]$ && $\frac{1}{28}(13F_8^{(3)}-76F_{16}^{(3)})$ && -1 && 0 && -1\\
\midrule
Total &$\phantom{aa}$&&$\phantom{aa}$&$\frac{5}{7}(8F_{16}^{(3)}-F_8^{(3)})$ &$\phantom{aa}$& $\frac{5}{16}(8F_{16}^{(3)}-F_8^{(3)})$&$\phantom{aa}$& $\frac{5}{7}(8F_{16}^{(3)}-F_8^{(3)})$ \\
\bottomrule
\end{tabular}
}
\end{center}
\caption{Sums of the terms appearing in $\Xi_8^{(3)}[0^{(3)}]$. In the third, fourth and fifth columns we report the coefficients of the $O^+$-invariants appearing in the expression of $\Xi_8^{(3)}[0^{(3)}]$ in the three basis.}
\label{tab:sum3}
\end{table}

\subsection{Genus four}
As for the genus three case, we report in Table \ref{tab:sum4} the sums of each term appearing in the form
$\Xi_8^{(4)}[0^{(4)}]$. Again, for the three equivalent bases of the space of $O^+$-invariants, the cosmological
constant vanishes on the Jacobi locus due to the vanishing of the form $J^{(4)}$. It should be noted that the cosmological constant vanishes just on the moduli space of curves even if the forms $\Xi_8^{(4)}[\Delta^{(4)}]$ are well defined on the whole $\HH_4$.
\begin{table}[h!]
\begin{center}
\resizebox*{1\textwidth}{!}{
\begin{tabular}{ccccccccc}
\toprule
Function && Sum && CDG && DP && Gr \\
\midrule
$F_1^{(4)}$ && $F_{16}^{(4)}$ && $\frac{1}{6}$ && $\frac{1}{6}$ && $\frac{1}{16}$\\
$F_2^{(4)}$ &&  $16F_{16}^{(4)}$ && $\frac{1}{3}$ && $\frac{1}{3}$ && 0\\
$F_3^{(4)}$ && $F_8^{(4)}$ && $-\frac{1}{8}$ && $-\frac{1}{8}$ && 0\\
$F_8^{(4)}$ && $136F_8^{(4)}$ && 0 &&$\frac{1}{64}$ && 0\\
$F_{88}^{(4)}$ && $32F_8^{(4)}-32F_{16}^{(4)}$ && 0 && $-\frac{1}{16}$ && 0\\
$F_{16}^{(4)}$ && $136F_{16}^{(4)}$ && 0 && 0 && 0 \\
$G_1^{(4)}[0^{(4)}]$ && $F_{8}^{(4)}-F_{16}^{(4)}$ && 0 && 0 && $-\frac{1}{16}$\\
$G_2^{(4)}[0^{(4)}]$ && $43F_{16}^{(4)}-\frac{1}{2}F_{8}^{(4)}$ && 0 && 0 &&$\frac{1}{8}$\\
$G_3^{(4)}[0^{(4)}]$ && $\frac{15}{7}(\frac{3}{4}F_8^{(4)}-5F_{16}^{(4)})$ && $-\frac{1}{2}$ && $-\frac{1}{2}$ && $-\frac{1}{2}$\\
$G_4^{(4)}[0^{(4)}]$ && $\frac{29}{7}F_{16}^{(4)}-\frac{11}{56}F_{8}^{(4)}$ && 4 && 0 && 4\\
\midrule
Total &$\phantom{aa}$&&$\phantom{aa}$& $\frac{12}{7}(16F_{16}^{(4)}-F_8^{(4)})$ &$\phantom{aa}$& $\frac{45}{56}(16F_{16}^{(4)}-F_8^{(4)})$ &$\phantom{aa}$& $\frac{12}{7}(16F_{16}^{(4)}-F_8^{(4)})$ \\
\bottomrule
\end{tabular}
}
\end{center}
\caption{Sums of the terms appearing in $\Xi_8^{(4)}[0^{(4)}]$. In the third, fourth and fifth columns we report the coefficients of the $O^+$-invariants appearing in the expression of $\Xi_8^{(4)}[0^{(4)}]$ in the three basis.}
\label{tab:sum4}
\end{table}

\subsection{Genus five}
As for the two previous cases we report in Table \ref{tab:sum5} the sums of each term appearing in the form
$\Xi_8^{(5)}[0^{(5)}]$. In the Table the functions
$G_i^{(5)}[0^{(5)}]$, $i=0,\cdots,5$, with $G_0^{(5)}[0^{(5)}]\equiv
F_1^{(5)}$, are the same as in \cite{Grr}. In the genus five case the
cosmological constant no longer vanishes neither on $J_5$. Actually,
it was shown in \cite{GS} that the zero locus of $J^{(5)}$ is the
locus of trigonal curves. Following \cite{OPSY}, if we subtract from the forms
$\Xi_8^{(5)}[0^{(5)}]$ the value of the cosmological constants divided
by 528, the number of the even characteristics in genus five, we
obtain again a function satisfying the three constraints and,
moreover, having zero cosmological constant. The correct
factorization is due to the fact that the form $J^{(5)}$ vanishes
when restrict both on $\HH_1\times\HH_4$ and on
$\HH_2\times\HH_3$. Moreover, this consideration shows that in genus
five the three constraints no longer assure the uniqueness of the form
$\Xi_8^{(5)}[0^{(5)}]$ because we can always add a multiple of the Schottky
relation that is not zero on $J_5$ obtaining another forms with the
correct behaviour.
%Again, for the three equivalent basis of the space of $O^+$-invariant, the cosmological
%constant vanishes due to the vanishing of the form $J^{(4)}$. It should be noted that the cosmological constant vanishes just on the moduli space of curves even if the forms $\Xi_8^{(4)}[\Delta^{(4)}]$ are well defined on the whole $\HH_4$.
\begin{table}[h!]
\begin{center}
\[
\begin{array}{ccccccc}
\toprule
\mbox{Function} && \mbox{Sum} && \mbox{DP} && \mbox{Gr} \\
\midrule
F_1^{(5)} && F_{16}^{(5)} && \frac{1}{12} && \frac{1}{32} \\
F_2^{(5)} &&  32F_{16}^{(5)} && \frac{1}{3} && 0 \\
F_3^{(5)} &&  F_8^{(5)} && -\frac{1}{16} && 0 \\
F_8^{(5)} && 528F_8^{(5)} && -\frac{1}{32} && 0  \\
F_{88}^{(5)} && 128 F_8^{(5)}-128 F_{16}^{(5)} && \frac 18 && 0 \\
F_{16}^{(5)} && 528F_{16}^{(5)} && 0 && 0 \\
G_1^{(5)}[0^{(5)}] && F_{8}^{(5)}-F_{16}^{(5)} && 0 && -\frac{1}{32} \\
G_2^{(5)}[0^{(5)}] && 171F_{16}^{(5)}-\frac{1}{2}F_{8}^{(5)} && 0 && \frac{1}{16} \\
G_3^{(5)}[0^{(5)}] && \frac{173}{28}F_8^{(5)}-\frac{299}{7}F_{16}^{(5)} && -\frac{1}{4} && -\frac{1}{4} \\
G_4^{(5)}[0^{(5)}] &&
\frac{389}{7}F_{16}^{(5)}-\frac{43}{56}F_{8}^{(5)} && 2 && 2 \\
G_5^{(5)}[0^{(5)}] &&
-\frac{733}{217}F_{16}^{(5)}+\frac{475}{3472}F_8^{(5)} && 0 && -32 \\
\midrule
\mbox{Total}&$\phantom{aa}$&&$\phantom{aa}$& \frac{51}{14}(32F_{16}^{(5)}-F_8^{(5)}) &$\phantom{aa}$& \frac{1632}{217}(32F_{16}^{(5)}-F_8^{(5)}) \\
\bottomrule
\end{array}
\]
\end{center}
\caption{Sums of the terms appearing in $\Xi_8^{(5)}[0^{(5)}]$. In the third, fourth and fifth columns we report the coefficients of the $O^+$-invariants appearing in the expression of $\Xi_8^{(5)}[0^{(5)}]$ in the two basis.}
\label{tab:sum5}
\end{table}

\subsubsection{Remark}
The sums reported in Table \ref{tab:sum3}, \ref{tab:sum4}, and \ref{tab:sum5} can be computed using a computer and a software
(for example,  Mathematica), and in any case follow directly from Lemma 9 in \cite{SM}.

\section{Conclusion}
In this work, enlightened by the representation theory of finite
groups, we have obtained some new expressions of the superstring measure at
 genus three and four in which the classical theta functions appear at higher
power than in previous works. For the genus five case we define a
function that on the Jacobi locus has the correct behaviour with
respect to the three constraints of Section \ref{strategy}. Although
this form satisfies all the constraints it is far to be unique. In
$g=5$ the constraints
are not strong enough to characterize it completely. Moreover, the non
normality of the ring of the modular forms in genus five and the
possibility of some (not known) relations between the theta constants
make it hard to
prove the uniqueness. At this point some more insight
in the physics leading  to the formulations of the  ans\"atze in
Section \ref{strategy} is necessary.
Some checks for the proposed ans\"atze for the measure are considered
by Morozov in \cite{Mo,Mo2}: he investigated the vanishing of
1,2,3-point functions. In \cite{GS2} Grushevsky and Salvati Manni
proved the vanishing of the 2-point function for $g=3$. 
Moreover, in \cite{MV} Matone and Volpato show how the chiral
superstring amplitudes can be obtained through factorisation of the
higher genus chiral measure. In \cite{MV2} they also discuss the
vanishing of the three-point amplitude at three loop.
These facts open some general questions about the formulation of string theory in
the perturbative approach. The ans\"atze of D'Hoker and Phong,
modified as in \cite{CDG}, don't suffice to characterise
the measure uniquely.
Moreover, is the general form of the amplitudes \eqref{ampl}, proposed by D'Hoker and Phong, from which one works out
the expression of the measures correct? Some questions arise about this last point and about the fibration that led
to \eqref{ampl}, see \cite{CD} for a discussion.
Again, the expression for the forms $\Xi_8^{(g)}[0^{(g)}]$ found in
\cite{OPSY} using the formalism of theta series open new questions about
the uniqueness of the superstring measure.
%and about the relations
%between the space of modular forms built with the classical theta
%functions and with the lattice theta series.

These points seem to reopen some old problems about the general formulation of the perturbative approach to superstring
theory.

\subsection*{Acknowledgements}
I am grateful to Sergio L. Cacciatori and Bert van Geemen for several interesting and stimulating discussions, for some advices on how the superstring measure could be rewritten and for detailed comments on the first versions of this paper. I like to thank Vittorio Gorini for suggestions.
I am also indebted with Riccardo Salvati Manni for explaining me some aspects about the dimensions of certain spaces of modular forms, for his comments on the draft and for a pleasant meeting in Milano.

\section{A: Moduli space and Schottky problem}\label{rs}
Here we briefly summarise some topics about the moduli space of Riemann surfaces, for details see \cite{vG2}.

Let $C$ be a Riemann surface of genus $g$ and consider the homology group $H_1(C,\ZZ)\cong \ZZ^{2g}$. A symplectic basis
of $H_1(C,\ZZ)$ is a basis $\{\alpha_1,\cdots,\alpha_g,\beta_1,\cdots,\beta_g\}$ satisfying
$(\alpha_i,\alpha_j)=(\beta_i,\beta_j)=0$ and $(\alpha_i,\beta_j)=\delta_{ij}$. Let $H^0(C,\Omega_C)$ be the
$g$-dimensional complex vector space of holomorphic one forms on $C$. Given a path $\gamma\in C$ and an
$\omega \in H^0(C,\Omega_C)$ one can compute the integral $\int_\gamma \omega$ and, if $\gamma$ is a closed path,
the integral depends only on the homology class of $\gamma$. It can be shown that given a symplectic basis for $H_1(C,\ZZ)$
then there is a unique basis $\{\omega_1,\cdots,\omega_g\}$ of $H^0(C,\Omega_C)$ such that
$\int_{\alpha_i}\omega_j=\delta_{ij}$. We now use the $\beta_j$ to define a complex $g\times g$ matrix, the period
matrix of $C$, $\tau=(\tau_{ij})\in M_g(\CC)$ with $\tau_{ij}:=\int_{\beta_i} \omega_j$, where $\omega_i$ is an element
of the basis of $H^0(C,\Omega_C)$ determined from the symplectic basis. Torelli's theorem asserts that one can recover
the Riemann surface from its period matrix.
The Schottky problem basically asks for equations which determine the period matrices of Riemann surfaces among all
$g\times g$ matrices. Period matrices have two properties: they are symmetric and $\Imm(\tau)$, the imaginary part of
$\tau$, which is a symmetric, real, $g\times g$ matrix, defines a positive definite quadratic form on
$\RR^g$: ${}^tx(\Imm \tau)x>0$, for all $x\in\RR^g$; briefly one writes $\Imm(\tau)>0$. This leads to the definition of
the Siegel upper half plane $\HH_g:=\{\tau\in M_g(\CC):\;{}^t\tau=\tau,\,\Imm(\tau)>0\}$. Thus if $\tau$ is the
period matrix of a Riemann surface, then $\tau\in \HH_g$.
One can show that $\HH_g$ is a complex manifold of dimension $\frac{1}{2}g(g+1)$.

To define the period matrix of a Riemann surface we had to choose a symplectic basis and two such basis are related by
an element of the symplectic group $\Gamma_g$. The symplectic group acts on $\HH_g$ and the period matrix of
Riemann surfaces are a $\Gamma_g$-orbit in $\HH_g$. Thus one can study the images of period matrices under the quotient
map $\pi:\HH_g\rightarrow A_g:=\Gamma_g\backslash \HH_g$.
The moduli space $M_g$ of Riemann surfaces is a variety whose points correspond to isomorphism classes of Riemann
surfaces. Then we have a well defined holomorphic map: $j:M_g\rightarrow A_g$, $[X]\mapsto \Gamma_g\tau$, where $\tau$
is a period matrix of $X$. This map is injective from Torelli's
theorem. The Schottky problem can now be reformulated as
the problem of finding equations for the image of $j$.

Let $J_g^0\subset\HH_g$ be the set of period matrices of Riemann surfaces. Its image in $A_g$ is
$j(M_g)=\mbox{Image}(J_g^0\rightarrow A_g=\Gamma_g\backslash
\HH_g)$. We have the diagram:
\be \nonumber
\xymatrix{
J_g^0 \ar @{^{(}->}[r]^{i} & \HH_g \ar[d]_\pi \\
M_g \ar[r]^(.3){j} & A_g:=\Gamma_g\backslash \HH_g
}
\ee
where $i$ is the immersion map of $J_g^0$ in $\HH_g$ and $J(M_g)=\pi(i(J_g^0))$.
The subvariety $J_g^0$ and $j(M_g)$ are not
closed and one defines the Jacobi locus $J_g$ as the closure of $J_g^0$ in $\HH_g$. A $\tau\in\HH_g$ will be
called decomposable if $\tau$ lies in the $\Gamma_g$-orbit of matrices in diagonal block form. The set $J_g-J_g^0$
in $\HH_g$ consists of decomposable matrices, the diagonal blocks being period matrices of Riemann surfaces of lower
genus.
From Teichm\"uller theory one knows that the subset $J_g$ is actually an irreducible subvariety of $\HH_g$ of
dimension $3g-3$, for $g>1$ and for $g=1$ one has $\HH_1=J_1=J_1^0$. The Table \ref{tab:dim} shows that the
Schottky problem is trivial for $g\leq 3$. This shows why for $g\leq 3$, as expected, the forms $\Xi_8^{(g)}[0^{(g)}]$
are defined on the whole $\HH_g$.

\end{document}